\documentclass[twocolumn]{aastex62}

\usepackage{longtable}
\usepackage{graphicx}
\usepackage{amsmath,amssymb}
\usepackage{color}
\usepackage{units}
\usepackage{epstopdf}
\usepackage{hyperref}
\usepackage{multirow}

\usepackage{subfigure}
\usepackage{booktabs}

\newcommand{\beq}{\begin{equation}}
\newcommand{\eeq}{\end{equation}}

\definecolor{Gray}{gray}{0.9}
\definecolor{orange}{rgb}{0.9,0.5,0}

\graphicspath{{./plots/}}

\begin{document}

\title{Constraints on the neutron star equation of state from AT2017gfo using radiative transfer simulations}

\author{Michael W. Coughlin}
\affil{Division of Physics, Math, and Astronomy, California Institute of Technology, Pasadena, CA 91125, USA}

\author{Tim Dietrich}
\affil{Nikhef, Science Park 105, 1098 XG Amsterdam, The Netherlands}

\author{Zoheyr Doctor}
\affil{Kavli Institute for Cosmological Physics, University of Chicago, Chicago, IL 60637, USA}
\affil{Department of Physics, University of Chicago, Chicago, Illinois 60637, USA}

\author{Daniel Kasen}
\affil{Departments of Physics and Astronomy, and Theoretical Astrophysics Center, University of California,
Berkeley, California 94720-7300, USA}
\affil{Nuclear Science Division, Lawrence Berkeley National Laboratory, Berkeley, California 94720-8169, USA}

\author{Scott Coughlin}
\affil{Physics and Astronomy, Cardiff University, Cardiff, CF10 2FH, UK}
\affil{Center for Interdisciplinary Exploration \& Research in 
(CIERA), Northwestern University, Evanston, IL 60208, USA}

\author{Anders Jerkstrand}
\affil{Max-Planck Institut f{\"u}r Astrophysik, Karl-Schwarzschild-Strasse 1, D-85748 Garching, Munich, Germany}

\author{Giorgos Leloudas}
\affil{Dark Cosmology Centre, Niels Bohr Institute, University of Copenhagen, Juliane Maries Vej 30,
2100 Copenhagen, Denmark}

\author{Owen McBrien}
\affil{Astrophysics Research Centre, School of Mathematics and Physics, Queen's University Belfast, Belfast BT7 1NN, Northern Ireland UK}

\author{Brian D.~Metzger}
\affil{Department of Physics and Columbia Astrophysics Laboratory, Columbia University, New York, New York 10027, USA}

\author{Richard O'Shaughnessy}
\affil{Center for Computational Relativity and Gravitation, Rochester Institute of Technology, Rochester, New York 14623, USA}

\author{Stephen J. Smartt}
\affil{Astrophysics Research Centre, School of Mathematics and Physics, 
       Queen's University Belfast, Belfast BT7 1NN, UK}
       
\begin{abstract}
The detection of the binary neutron star merger GW170817 together with the observation of electromagnetic counterparts across the entire spectrum 
inaugurated a new era of multi-messenger astronomy.
In this study we incorporate 
wavelength-dependent opacities and emissivities calculated from 
atomic-structure data enabling us to model
both the measured lightcurves and spectra
of the electromagnetic transient AT2017gfo.
Best-fits of the observational data are obtained by 
Gaussian Process Regression, which allows us to present 
posterior samples for the kilonova 
and source properties connected to GW170817. 
Incorporating constraints obtained from the gravitational wave signal
measured by the LIGO-Virgo Scientific Collaboration,  
we present a $90\%$ upper bound on the mass ratio 
$q \lesssim 1.38$ and a lower bound on the tidal 
deformability of $\tilde{\Lambda} \gtrsim 197$, 
which rules out sufficiently soft equations of state. 
Our analysis is a path-finder for more realistic kilonova models
and shows how the combination of gravitational wave and electromagnetic measurements 
allow for stringent constraints on the source parameters and 
the supranuclear equation of state.
\end{abstract}


\section*{Introduction}

A new era of multi-messenger astronomy began with the combined detection of a neutron star (NS) merger via the gravitational wave GW170817 \citep{AbEA2017b}, 
the gamma-ray burst (GRB) GRB170817A \citep{AbEA2017e}, and the electromagnetic (EM) transient 
AT2017gfo \citep{AbEA2017g}. The discovery of a bright optical and near-infrared source in NGC4993, consistent with the gravitational-wave sky localization, during the first 12\,hrs after the joint gravitational wave and gamma ray detections  
\citep{ArHo2017,2017Sci...358.1556C,LiGo2017,SoHo2017,TaLe2017,VaSa2017} led to intensive follow-up campaigns to show that this was an unusual and unprecedented transient emitting from the X-ray to radio
\citep{2017ApJ...848L..21A,ChBe2017,CoBe2017,2017Sci...358.1570D,2017Sci...358.1565E,2017ApJ...848L..25H,2017Sci...358.1579H,KaNa2017,KiFo2017,2017ApJ...848L..20M,2017ApJ...848L..32M,NiBe2017,2017Sci...358.1574S,2017Natur.551...67P,SmCh2017,2017Natur.551...71T,2017PASJ...69..101U}.
This event showed that compact binary mergers including at least one NS 
can create an EM counterpart known as a {\it kilonova} 
\citep{LaSc1974,LiPa1998,MeMa2010,RoKa2011,KaMe2017}. 
Kilonovae originate from neutron-rich outflows from the merger which emit ultra-violet/optical/infrared 
emission powered by the radioactive decay of r-process elements.  Kilonovae are of enormous scientific value: They offer insight into 
the equation of state (EOS) of NSs \citep{BaBa2013,AbEA2017b,RaPe2018,BaJu2017}, 
the formation of heavy elements \citep{JuBa2015,WuFe2016,RoLi2017,AbEA2017f},
and the expansion rate of the universe \citep{2017Natur.551...85A}.

While AT2017gfo is the only confirmed kilonova observed to date, there has been significant theoretical work on modeling the nature of these transients.  These studies have postulated two main forms of ejecta from NS mergers: dynamical and wind ejecta.  The dynamical ejecta is the matter expelled at the moment of the merger from tidal stripping of the NSs and from the NS-NS contact interface \citep[e.g.][]{RoLi1999,OeJa2007,BaGo2013,WaSe2014,SeKi2015,HoWa2016,RoFe2017,WoKo2017}.  Wind ejecta is produced through remnant accretion disk winds, which can be driven by neutrino energy, magnetic fields, 
viscous evolution and/or nuclear recombination energy \citep[e.g.][]{FrWo1999,DiPe2002,MePi2008,DeOt2008,FeMe2013,PeRo2014,SiRi2014,JuBa2015,ReKu2014,CiSi2015,MaSt2015}.  
The masses, velocities, and compositions of the different ejecta types can vary, which
results in different observed kilonova morphology.

The UV - optical - near infrared lightcurves and spectra of 
AT2017gfo have been used to infer ejecta mass, velocities and compositions when 
combined with simple toy model approaches 
\citep[e.g. those of][]{1982ApJ...253..785A,2017LRR....20....3M}
and more sophisticated modelling of the few existing kilonova simulations, 
e.g.~\citep{KaMe2017}. 
The first papers published after the event which included quantitative modelling  
\citep{CoBe2017,KiFo2017,SmCh2017,TaLe2017} and later papers based on 
combined data \citep{ViGu2017,2017arXiv171005445R,PeRa2017,2017arXiv171109638W} 
produced broadly similar results. All the analyses consistently found that a few hundredths of a solar mass was ejected in AT2017gfo at velocities between 
0.1 - 0.3 c. However, none of these studies performed fits or inference using full radiative-transfer simulations.

In this work, we build on these previous analyses by performing Bayesian inference on observed AT2017gfo photometry and spectra using ``surrogate" models that are trained on the outputs of radiative transfer simulations. The surrogate models allow one to calculate the likelihood of the data for any ejecta parameters and hence derive posterior distributions on those parameters.  Additionally, we go beyond inferences of only ejecta properties and constrain the NS-binary parameters information from 
full numerical relativity simulations of NS mergers. 
The contribution of each type of ejecta and their mass, velocity, and composition are expected to depend on the parameters of the compact binary, the compact object masses, spins, orbital eccentricity, as well as the properties of NSs, such as the EOS \citep{RoLi1999,BaGo2013,HoKi13,LeLi2016,RaGa2016,DiUj2017,SiMe2017,AbEA2017f}. As such, observed kilonova emission can be used to constrain the compact binary parameters (or vice versa) using a mapping from ejecta properties to NS-binary parameters \citep{CoDi2017,AbEA2017c}. Of particular interest is the EOS of cold supranuclear matter, since it was constrained by the GW170817 signal \citep{AbEA2017b} and can be independently constrained by the electromagnetic data~\citep{RaPe2018}.

The layout of this paper is as follows: 
First, we describe the dataset used for our analysis. 
Then, we discuss our method for interpolating the output of 
kilonova simulations over the full parameter space of ejecta mass, velocity, and composition
and describe the Bayesian procedure for inferring ejecta properties of AT2017gfo 
from the photometry. 
Finally, we use the measured ejecta properties to put new constraints 
on the NS EOS and the GW170817 binary mass ratio.   

\section*{Data}
\label{sec:data}

A massive photometric data set was gathered with intra-day time resolution by many teams with latitudinally separated observatories in the southern hemisphere and in Hawaii. We compiled our own selected set of photometry and recalculated bolometric luminosities with realistic error bars. 
We initially took the photometry from the UV to $K-$band from 
\citep{2017PASA...34...69A,ArHo2017,ChBe2017,CoBe2017,2017Sci...358.1570D,2017Sci...358.1565E,KaNa2017,TaLe2017,2017Natur.551...67P,2017Natur.551...71T,SmCh2017,2017PASJ...69..101U,VaSa2017} 
from phases +0.467d to +25.19d after GW170817 and at each epoch created the broadest 
spectral energy distribution possible. Data from the Swift satellite in UV bands were only 
available from \cite{2017Sci...358.1565E} until +1\,d and the last $U$-band detection is 
from \cite{SmCh2017} at +1.505\,d. No secure optical data are available after epoch
+11.3\,d when AT2017gfo faded below 24\,mag in $g$-band, and the transient is only detected in 
$H$ and $K_{\rm s}$ until +14.3\,d and then only $K_{\rm s}$ thereafter. 

We began with the photometry of \cite{SmCh2017} as the core data set and employed difference imaging at all epochs of PESSTO \citep[Public ESO Spectrosopic Survey of Transient Objects; ][]{2015A&A...579A..40S}, GROND and Pan-STARRS imaging.   
Our approach was to:  
i) complement this photometry only when this was necessary either due to insufficient temporal or wavelength coverage
ii) primarily use only $grizyJHK_{\rm S}$ AB mag photometry from sources that used image subtraction \citep[mostly DECam and Skymapper]{CoBe2017,2017PASA...34...69A}, or from HST where host contamination is not important \citep{TaLe2017}
iii) when this was not possible, focus on a small number of independent sources such as Gemini South \citep{KaNa2017}, VISTA \citep{TaLe2017} and Sirius \citep{2017PASJ...69..101U}. We verified consistency between the data sets through direct comparison.  In this way, we 
compiled $grizyJHK_{\rm S}$ SEDs, or as broad a subset as the data allowed. 
From the first detection at 0.47\,d, there are five distinct epochs within the first 24\,hrs (including Swift satellite data) at which $L_{\rm bol}$ can be calculated. A total of 20 distinct epochs with enough data to define a black body fit can be defined up to +10.4\,d after GW170817. 
We note that our GROND $K-$band photometry has been updated compared to \cite{SmCh2017}. This is 
because the GROND template for host subtraction still contained flux from the transient
\citep[as first noted by][]{ViGu2017}. The image subtraction has now been redone using 
a different template with no flux present and 
after this correction, the present GROND light-curve matches much better with other $K-$band measurements in the literature. The recommended updated photometry values are now published and available on the PESSTO webpage\footnote{www.pessto.org} and we employ them here. We used this 
$ugrizyJHK_{\rm S}$ compilation to constrain the model fits as discussed below. 

We have
used these data to calculate the bolometric luminosities from +0.467\,d to +13.21\,d\footnote{We use the data up to 10\,d when calculating the fits.}, after
which the wavelength coverage is insufficient to securely determine $L_{\rm bol}$. The bolometric lightcurves are given in Table~\ref{table:bolometric} and their construction in Appendix~\ref{sec:lightcurves}.
Manual comparison of the models of \cite{KaMe2017} showed some promising agreement with the near infra-red spectrum of \cite{ChBe2017} at +2.5\,d to +4.5\,d in particular, although only the 
1.0-1.8$\mu$m region was compared and the evolution was not consistently reproduced. 
It is clear that the X-shooter spectra 
of \cite{PiDa2017} and \cite{SmCh2017} 
taken with ESO's Very Large Telescope contain all available spectral information since they
cover 0.35-2.5$\mu$m on a daily basis from +1.5\,d to +10.5\,d. 
This is an excellent dataset to more rigorously constrain the ejecta properties. 
We employed the reduced X-Shooter spectra made publicly available on WISeREP\footnote{https://wiserep.weizmann.ac.il} and through PESSTO$^{1}$. We do not use any other spectral data set, as other data is either 
inferior signal-to-noise, reduced wavelength coverage, or both, and after +1.5\,d, no other spectral dataset provides additional temporal information that enhances the X-shooter sequence in any way. 

\section*{Kilonova Surrogate Model}

Throughout this work, we use the kilonova models presented in \cite{KaMe2017} which 
employ a multi-dimensional Monte Carlo code to solve the multi-wavelength radiation transport 
equation for a relativistically expanding medium.  Initial use of the model and comparison
to data showed  promising similarities with some epochs of near infra-red spectra 
\citep{ChBe2017} and the bolometric luminosity \citep{KiFo2017}. 
Until now a comparison with the full wavelength and temporal spectral series 
\citep[X-Shooter spectra from][]{PiDa2017,SmCh2017} has not been done, but  
is essential to extract additional details about the ejecta \citep{SmCh2017,2017arXiv171005445R,2017arXiv171109638W}. 
Here we will employ all of the data
published to date to constrain the model fits. 
 
The \cite{KaMe2017} models depend parametrically on the ejecta mass 
$M_{\rm ej}$, the mass fraction of lanthanides 
$X_{\rm lan}$, and the ejecta velocity $v_{\rm ej}$.
In terms of the underlying physics of the merger and ejecta processes described above, these three parameters would be determined by the detailed ejecta processes involved e.g. the duration of the outflow, mass involved, and nucleosynthesis allowed, given the outflow trajectory and neutrino illumination sources. In this work, eschewing detailed neutrino radiation hydrodynamics simulations of mergers, we treat these properties as parameters.
We can use separate 1-component models to create a 2-component ejecta model by summing 
together two 1-component models. 
This sum is performed by first generating the bolometric lightcurves, photometric lightcurves, and spectra for the individual models. 
The 2-component bolometric lightcurves and spectra are produced by simply adding the 1-d curves together, while the photometric lightcurves are added in the way appropriate for log-based quantities.
The use of a 2-component model is motivated by 
both the theoretical prediction of the presence 
of different ejecta components and also by the fact that the ejecta are observed to fade 
rapidly in the UV and optical but have a significantly different near-infrared evolution. 
We restrict our analysis to spherical symmetry and a uniform composition, 
and neglect mixing of different ejecta types \citep{RoFe2017}
when we add the 2 separate model components.
The expansion of the model to non-spherical geometries and 
compositional gradients is left for future analyses.

The model provided in \cite{KaMe2017} and described above is produced on a grid with ejecta 
masses 
$M_{\rm ej} [M_\odot]$ = 0.001, 0.0025, 0.005, 0.0075, 0.01, 0.25, 0.05, and 0.1, ejecta velocities 
$v_{\rm ej} [c]$ = 0.03, 0.05, 0.1, 0.2, and 0.3, and mass fraction of lanthanides 
$X_{\rm lan}$ = 0, $10^{-5}$, $10^{-4}$, $10^{-3}$, 
$10^{-2}$, and $10^{-1}$.
The models have temporal epochs of 0.1\,day sampling.
In order to draw inferences about generic sources not corresponding to one of these gridpoints, we develop a novel method to create a parameterized 
model from a set of numerical data.
We adapt the approach outlined in \cite{DoFa2017} and \cite{Purrer2014}, 
where Gaussian Process Regression (GPR) is employed to interpolate principal components of gravitational waveforms 
based on existing sets of simulations. 
In this analysis, we perform a similar computation 
but on bolometric luminosities, 
lightcurves in standard filters, and spectra. The details of the algorithm to perform the interpolation can be found in Appendix~\ref{sec:surrogate}.
We also explore in Appendix~\ref{sec:fits} the question of whether there are enough simulations on the grid in order to draw inferences based on the model. We show by removing a simulation from the grid and comparing the resulting interpolated lightcurves and spectra to that simulation that the grid is dense enough to reproduce the simulation.

\section*{Analysis}

\begin{table*}[t]
\caption{Ejecta properties estimated from the GPR. 
The estimated uncertainties give the $1\sigma$-uncertainty. Corner plots from which these numbers are derived are shown in Appendix~\ref{sec:corner}. The 2 component model lists the higher lanthanide fraction as $X_{\rm lan1}$ and lower as $X_{\rm lan2}$ (corresponding to dynamical and wind components).
}
\label{tab:ejecta estimates} 
\setlength{\tabcolsep}{9pt}
\centering
\begin{tabular}{l|cc|cc|cc}
\toprule
                    & \multicolumn{2}{c|}{bolometric luminosity}    & \multicolumn{2}{c|}{lightcurve} & \multicolumn{2}{c}{spectra}\\ 
                    & 1 component           &       2 component     &  1 component & 2 component     & 1 component & 2 component \\ 
\hline
$\log_{10}(M_{\rm ej1}/M_\odot)$ &$-1.39^{+0.13}_{-0.11}$&$-2.50^{+1.06}_{-1.60}$&$-1.30^{+0.10}_{-0.13}$&$-1.51^{+0.23}_{-0.27}$&$-1.48^{+0.13}_{-0.14}$&$-2.03^{+0.56}_{-1.02}$\\
$v_{\rm ej1}$ [$c$]       &$+0.12^{+0.09}_{-0.06}$&$+0.09^{+0.09}_{-0.06}$&$+0.23^{+0.06}_{-0.16}$&$+0.10^{+0.08}_{-0.06}$&$+0.20^{+0.003}_{-0.004}$&$+0.10^{+0.08}_{-0.05}$\\
$X_{\rm lan1}$      &$-6.77^{+1.80}_{-1.30}$&$-2.18^{+1.56}_{-1.16}$&$-3.54^{+0.39}_{-0.36}$&$-1.61^{+0.96}_{-1.04}$&$-2.97^{+0.30}_{-0.39}$&$-1.52^{+0.97}_{-0.98}$\\
$\log_{10}(M_{\rm ej2}/M_\odot)$ &       --              &$-1.39^{+0.13}_{-0.63}$&     --                &$-1.59^{+0.16}_{-0.18}$&     --                &$-1.63^{+0.20}_{-0.34}$\\
$v_{\rm ej2}$ [$c$]      &       --              &$+0.20^{+0.05}_{-0.08}$&     --                &$+0.17^{+0.09}_{-0.10}$&     --                &$+0.20^{+0.03}_{-0.01}$\\
$X_{\rm lan2}$      &       --              &$-3.91^{+0.73}_{-0.72}$&     --                &$-4.73^{+0.41}_{-0.20}$&     --                &$-3.31^{+0.50}_{-0.77}$\\
\hline
\end{tabular} 
\end{table*}

\begin{figure}[t]
 \includegraphics[width=3.5in]{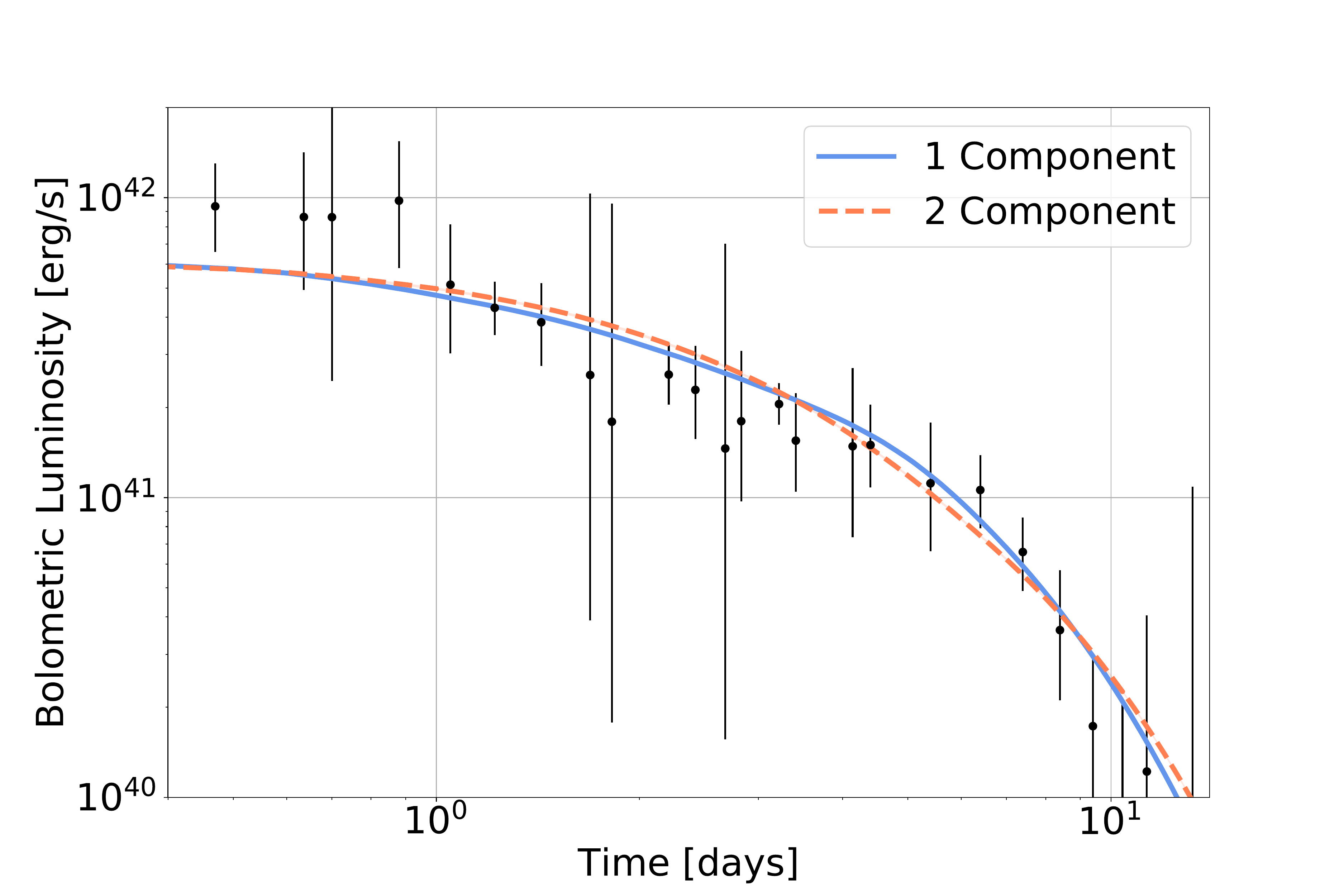}
  \caption{
   Derived bolometric luminosity and a maximum likelihood $\chi^2$ 
   fit using the 1- and 2-component kilonova bolometric luminosity models of \cite{KaMe2017}. We provide the $L_{bol}$ data in Table~\ref{table:bolometric}.
 }
 \label{fig:lbol}
\end{figure}

We use the Bayesian procedure described in  \cite{CoDi2017} to compare our GPR-based kilonova bolometric, photometric, and spectral models with the full observational data set and draw posterior inferences about our model parameters $v_{\rm ej},M_{\rm ej}$, and $X_{\rm lan}$. 
For each component, the flat priors used in our analysis cover the region
$-5 \leq \log_{10} (M_{\rm ej}/M_\odot) \leq 0$, $ 0 \leq v_{\rm ej} \leq 0.3$\,$c$, 
and $-9 \leq \log_{10} (X_{\rm lan}) \leq -1$.
In all cases, the likelihood is based on the $\chi^2$ value between our model and the data.
For the 2-component models, we require $X_{\rm lan 1}>X_{\rm lan 2}$ and $v_{1}<v_{2}$.  
The velocity prior is employed to limit to systems where the blue ejecta is ahead of the red ejecta, which is the regime for this non-interacting model to be valid.
The order of the components does reflect their lanthanide fraction, with a large $X_{\rm lan}$ corresponding to a red, 
lanthanide-rich component and a small $X_{\rm lan}$ to a blue, lanthanide-poor component. 
In fact, in the one-dimensional picture that we consider here, the blue component cannot be at lower velocity than the red physically because the latter would not allow its emission to escape. 

We now discuss this prior choice and the origin of the blue and red component of 
the kilonovae.
In general, there are two options. The first is that the ejecta is to a reasonable approximation isotropic, with a blue component everywhere ahead and faster
than the red one.
In this case, the present treatment of the multi-component model is appropriate, and all the conclusions derived are consistent.
There are reasons to expect this may be the case in certain regimes.
First proposed by \cite{MeFe2014}, it was thought the only source of the blue ejecta was from the disk wind in the case of a long-lived hypermassive NS and the red ejecta might arise from the tidal tail or a disk wind. The early spectral observations \citep{2017ApJ...848L..32M,NiBe2017,2017Sci...358.1574S,SmCh2017} suggest the blue component is moving relatively fast ($\approx 0.3\,$c) which is likely faster than a standard disk wind would produce, motivating its potential association with dynamical ejecta. This motivates our prior choice. 

There is also the possibility that the ejecta is significantly anisotropic or there are significant interactions between different components or with a possible expanding jet.
In general, a 2-component model where the components are allowed to interact would be required in this case, although the assumption above is valid in the case that the ejecta is observed from a specific direction such that the lanthanide-free component is ahead of and faster than the red one. The velocity constraints will not be valid if the red and blue components originate from geometrically distinct regions, e.g. if the blue comes out in the polar direction and the red comes out in the equatorial plane. For example, it has been shown that the polar dynamical ejecta could itself be blue \citep{WaSe2014,SeKi2015}. In addition, no numerical relativity simulations have produced ejecta masses seen from AT2017gfo ($\approx 0.05 M_\odot$) in the tidal tail component, while this quantity of red ejecta can readily come from the disk wind in the case that the hypermassive NS is relatively short-lived \citep{SiMe2017}. 
Recently, \cite{KaSh2018} used 2D radiative transfer models to show that the potentially anisotropic properties of the ejecta requires less dynamical and Lanthanide-free ejecta to reproduce AT2017gfo, reducing the tension with numerical relativity simulations.
Qualitatively similar results were seen in other studies using 2D models \citep{WoKo2017}, and in semi-analytical models that explicitly take into account the non-spherical character of the ejecta \citep{PeRa2017}.
Another possibility is a 2-component disk wind, e.g.~\citep{ShFu2017}. 
In this case, a fast, blue component is found for the 
outer torus ejection, and a slow red component for the inner.
For this reason, the results derived in the following rely on the assumption that the blue component is everywhere ahead and faster
than the red one, which may not be the case.

To validate our analysis procedure, we first reproduce previous bolometric and photometric analyses of this event. 
The first test is to reproduce the analysis in \cite{SmCh2017}, 
where the bolometric lightcurves were computed from the available photometry at that time. We fit our bolometric models to the bolometric data from \cite{SmCh2017} using a $\chi^2$ likelihood.
As shown in Figure~\ref{fig:lbol}, both the 1-component and the 2-component model 
can reproduce the measured bolometric luminosity. 
Although within error bars, the predicted bolometric luminosities are systematically low at early times.
Based on the 1-component fit to the bolometric luminosity, 
we estimate $\log_{10}(M_{\rm ej})  = -1.39$ ($M_{\rm ej}  = 0.041 M_\odot$), 
with a velocity of $v_{\rm ej}= 0.14$\,$c$ and a mass fraction of lanthanides of $X_{\rm lan} = 10^{-6.41}$
(see Table\,\ref{tab:ejecta estimates} for error bars and Appendix~\ref{sec:corner} for the associated corner plots).
Overall, this is consistent with \cite{SmCh2017} who found similar ejecta masses and velocities for 
a composition with an effective gray opacity of $\kappa \sim 0.1$\,cm$^2$/g. Uncertainties in the atomic data
render the conversion between opacity and lanthanide mass fraction non-trivial. However 
previous studies have shown that at $X_{\rm lan} \sim 10^{-1}$
models have an effective gray opacity of $\kappa \sim 10$, while $X_{\rm lan} \leq 10^{-6}$ models have an opacity closer to $\kappa \sim 0.1$, with the dependence being roughly logarithmic ($\kappa \propto [\log X_{\rm lan}]^{\alpha}$).
Employing a 2-component model fit to $L_{\rm bol}$ makes 
a consistent prediction for the light curve and results in a 
total ejected mass of $M_{\rm ej}  = 0.054 M_\odot$.  
While we can measure the total amount of ejecta by using only the bolometric information,
the amount of matter in each component (and their composition) is ill-determined; 
see the top row of the corner plots in Appendix~\ref{sec:corner}.\\

\begin{figure}[t]
 \includegraphics[width=3.5in]{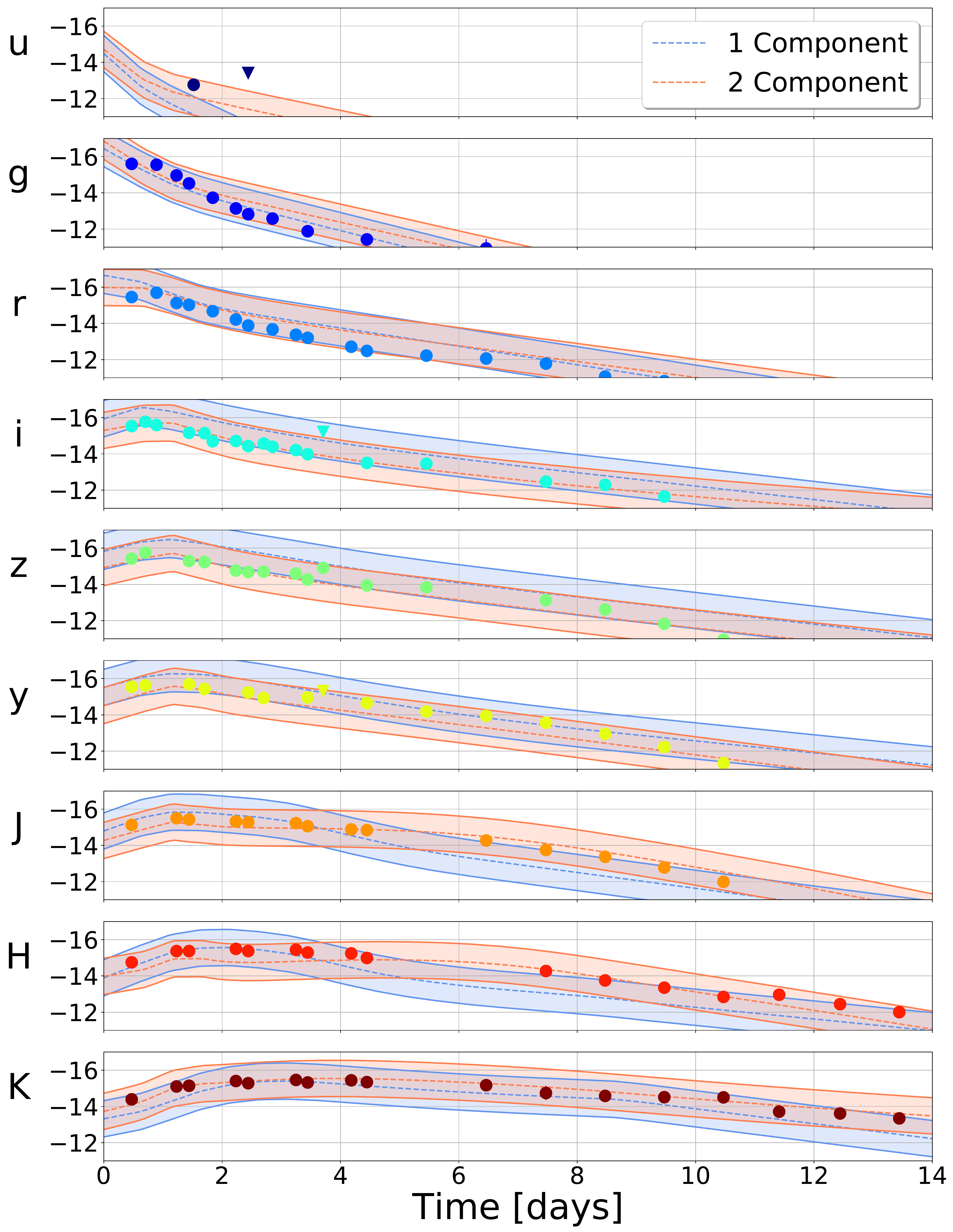}
  \caption{
  Lightcurves for both one and two component models from \cite{KaMe2017}.
   The shown lightcurves correspond to a maximum likelihood $\chi^2$ fit to the data.
   Shaded regions represent the assumed 1\,mag error budget. 
   The source of the photometry is summarized in Section\,\ref{sec:data}. 
 }
 \label{fig:lightcurve}
\end{figure}

Increasing the complexity of the analyzed data, 
we fit the broad band photometry points 
described earlier and illustrated in Figure~\ref{fig:lightcurve}. 
We assign model uncertainties of 1\,mag added in quadrature with 
the statistical error in the measured photometry \citep{CoDi2017}.
In general, the 1\,mag uncertainties, which are treated as 1-$\sigma$ errors, are designed to capture difficult-to-quantify systematic uncertainties, such as those in the the electron fraction and heating rate, which can lead to significant differences in the predicted luminosities \citep{RoFe2017}.
Fitting the lightcurves with a single component results 
in $\log_{10}(M_{\rm ej}/M_\odot)  = -1.41$ ($M_{\rm ej}  = 0.040 M_\odot$), consistent with our previous findings. 
However, for early times ($< 4$ days) 
the model does not allow a representation of the $H$-, and $K$-bands 
and the predicted $g$-band is not consistent within the assigned 
uncertainties after 4 days. 
Conversely, a 2-component model 
(blue shaded region) can reproduce both early and late-time behavior in all bands. 
Using photometric data, we can distinguish between the two types of ejecta 
with different velocities and lanthanide fractions.  These two components are not strongly differentiated using bolometric information alone. 
In our 2-component photometric analysis, we find that the more massive ejecta component 
has a higher lanthanide fraction.
The amount of blue (lanthanide-poor) ejecta is also notable, $\log_{10}(M_{\rm ej}/M_\odot)  = -1.59$ ($M_{\rm ej}  = 0.026 M_\odot$), forming a significant fraction of the total ejecta. We return to the implications for this in the summary.\\

\begin{figure}[t]
 \includegraphics[width=3.5in]{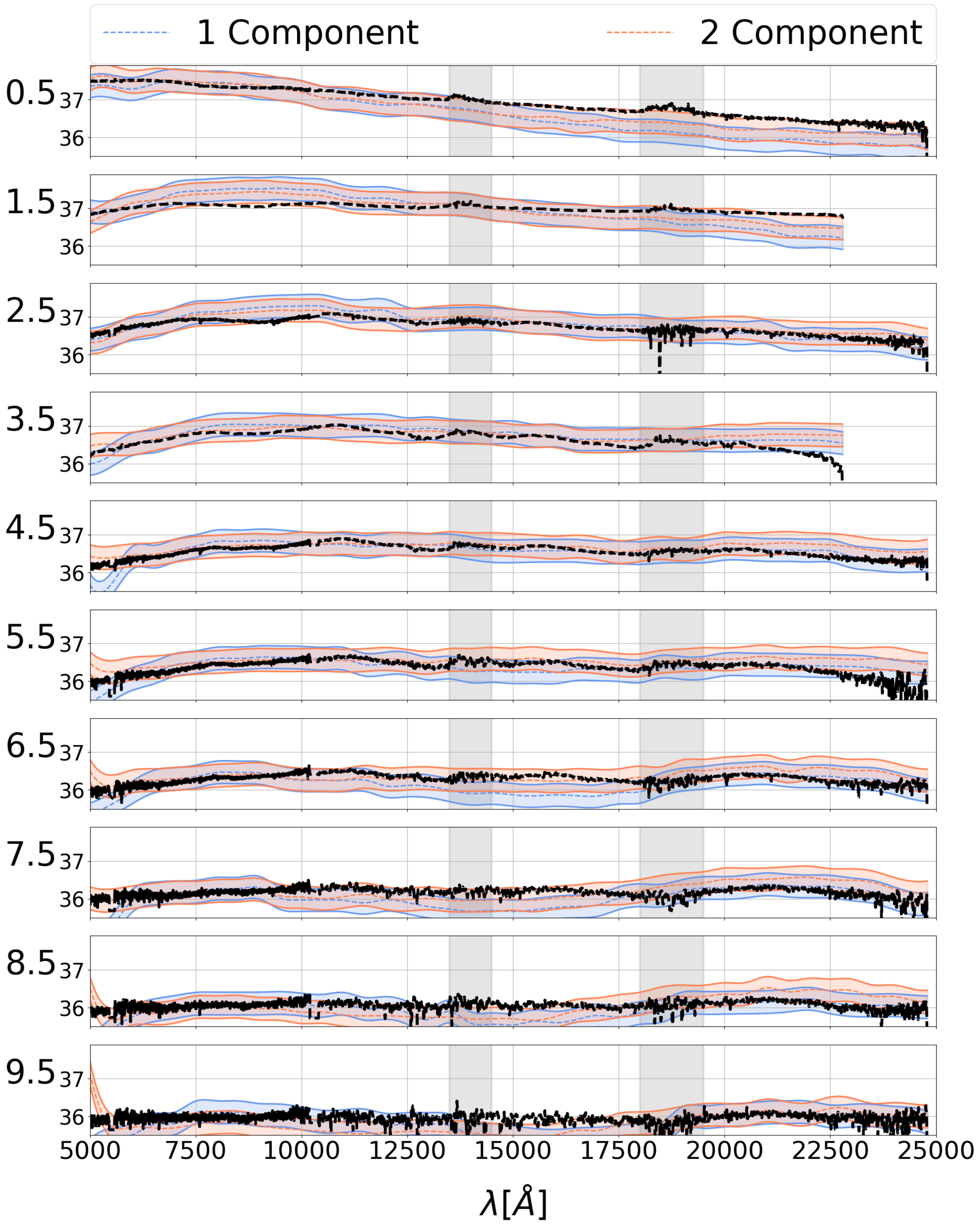}
 \caption{
   X-shooter spectra (black lines) in units of log10(ergs/s/A) at the available epochs (in units of days on the far left) and one and two component 
   model fits to the spectra \citep{PiDa2017,SmCh2017}.
   The shown spectra correspond to a maximum likelihood $\chi^2$ fit to the data.
   Shaded regions correspond to an assumed 1\,mag error budget. 
   The gray shaded regions mark ignored regions due to atmospheric transmission.
   }
  \label{fig:spectra}
\end{figure}

For the first time, we will also compare the spectra 
of AT2017gfo against theoretical kilonova predictions to compute posteriors. 
As discussed in \cite{PiDa2017} and \cite{SmCh2017}, the first X-Shooter and PESSTO EFOSC2 spectra are bright and blue, 
with rapid cooling just a day later. 
We fit the spectra of AT2017gfo directly \citep{PiDa2017,SmCh2017} in
figure~\ref{fig:spectra}. 
In line with the uncertainties of the photometric lightcurves, we use an upper error bar of 2.5$\times$ the spectral value, and a lower error bar of 1/2.5$\times$ the spectral value.
This model uncertainty is added in quadrature with the statistical error in the measured spectra.
Except for the early epoch when the predicted spectra declines slightly 
too quickly in the red, broad agreement in the overall shape between the kilonova model 
and the X-shooter spectra is obtained. 
Indeed, the model reproduces the 
spectra within the estimated uncertainty. 
The fit to the spectra results in 
$\log_{10}(M_{\rm ej}/M_\odot)  = -1.48$ ($M_{\rm ej}  = 0.033 M_\odot$) for a single component, 
and 
$\log_{10}(M_{\rm ej1}/M_\odot) = -2.03$ ($M_{\rm ej}  = 0.010 M_\odot$),  
$\log_{10}(M_{\rm ej2}/M_\odot) = -1.63$ ($M_{\rm ej}  = 0.023 M_\odot$) for the two component model. 
Overall, we find that the ejecta properties based on the lightcurves
and based on the spectra are very similar.
This shows that at the level of model uncertainties considered here, for a successful kilonovae 
model, it is possible to use either the lightcurves 
or the spectra, but the integrated information of the 
bolometric luminosity are insufficiently informative to constrain ejecta properties. 
We show in Appendix~\ref{sec:fits} that spectra based on the lightcurve fits (and vice-versa) give reasonable fits as well.

\section*{Inferring source properties}

\begin{figure}[t]
 \includegraphics[width=3.5in]{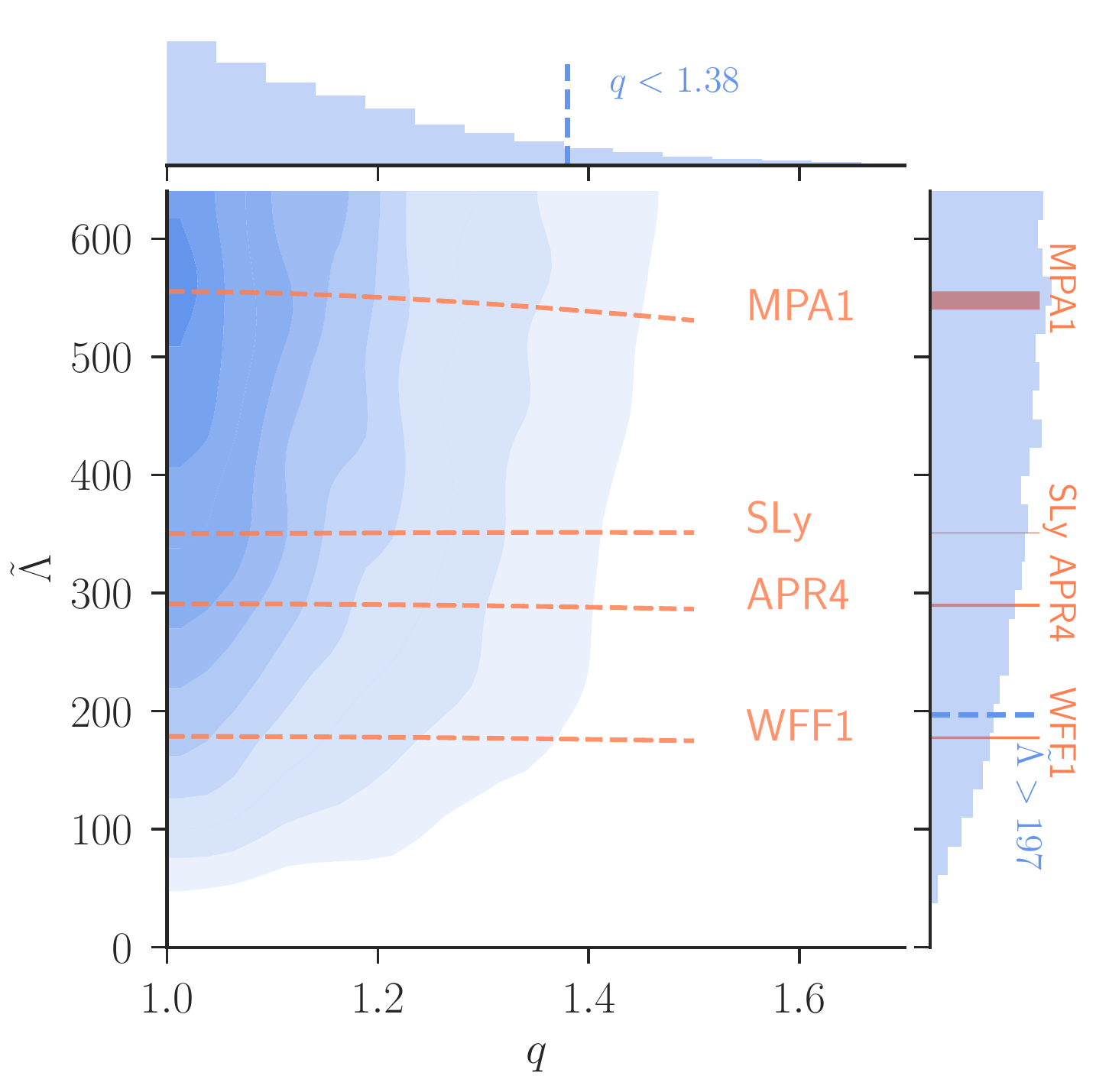} 
  \caption{
Corner plot for the constraining the mass ratio $q$, 
and tidal deformability $\tilde{\Lambda}$ assuming a chirp mass of 
$M_c=1.188M_{\odot}$ and based on the ejecta estimated obtained from the 
lightcurve fitting. 
We include estimates for the tidal deformability 
for a set of possible EOSs as orange lines showing that too soft 
EOSs are ruled out by our analysis.
The numbers represent the 90\% limits on the parameters.} 
\label{fig:BNS_properties}
\end{figure}

Finally, we want to use our analysis to obtain information 
about the binary parameters, such as the total mass, mass ratio, 
and tidal deformability. 
The idea follows the discussion in \cite{CoDi2017}: namely 
that information about the ejecta properties  
can be translated to constraints on the system parameters 
by fits such as those from \cite{DiUj2017}.
In this work, we improve on the fit of \cite{DiUj2017}, which connects 
the intrinsic binary parameters with dynamical ejecta properties extracted 
from full 3D numerical relativity simulations. 
These new fits are described in Appendix~\ref{sec:NR}.
We emphasize that numerical relativity simulations do not extend significantly past the moment of merger, and so they cannot capture the wind-driven ejecta expected at later times. We therefore for this study assume that the total ejecta mass is parameterized by the total ejected mass given by numerical relativity simulations with a scale factor such that 
\begin{equation}
M_{\rm ej} = A \times M_{\rm ej}^{\rm NR} \qquad \text{with}\ A > 1.  \label{eq:AMej}
 \end{equation}
We sample uniform in $A$ with broad 
enough priors so as to not affect 
the posteriors such that we only restrict 
$A \times M_{\rm ej}^{\rm NR}$ to be less than the total mass.

This fit allows us to directly tie the measured ejecta mass and velocity to properties of the binary, including the mass ratio and equation of state.
Based on this fit and the numerical relativity simulations that underly it, 
the total amount of dynamical ejecta will be largest when the NS involved are less compact.  Therefore, based on our estimates for the total amount of ejecta required to explain the kilonova as reported in  Table \ref{tab:ejecta estimates}, we expect that a self-consistent analysis of EM and GW data will disfavor NSs that are too compact and hence allow us to constrain the nuclear equation of state.

Incorporating information from gravitational-wave parameter estimation, namely a 
chirp mass $M_c$ of $M_c=1.188M_{\odot}$~\citep{AbEA2017b} and an upper limit on 
the tidal deformability of 
$\tilde{\Lambda} \lesssim$ 640~\footnote{The exact value of $\tilde{\Lambda} \lesssim 640$ arises from the fact that as stated in \cite{AbEA2017b}
an analysis of GW170817 with the \mbox{SEOBNRv4\_ROM\_NRtidal} waveform model~\cite{BoEA2016,DiBe2017,DiEA2018} gives an 80\% tighter bound than the PN based TaylorF2 model for which $\tilde{\Lambda}=800$ was stated.} we are able to place constraints on 
the mass ratio and tidal deformability of the system. Fig.~\ref{fig:BNS_properties} summarizes our findings. 
We find that the mass ratio of GW170817 is with 90\% 
confidence smaller than $q\lesssim 1.38$, while the 90\% lower bound
on the tidal deformability is $\tilde{\Lambda} \gtrsim 197$.   
This lower bound shows that more compact EOSs such as WFF1 are disfavored, 
see Fig.~\ref{fig:BNS_properties}.
These results can be compared to estimates obtained from a reanalysis of GW170817 \citep{DeFi18}, which incorporates quasi-universal relations for the tidal deformability and obtains 90\% lower bounds on the tidal deformability $\tilde{\Lambda} \gtrsim 117$ and 90\% upper bounds on the mass ratio $q \lesssim 1.51$.
Our analysis shows that even without the use of quasi-universal relations tighter constraints on the binary parameters can be obtained from EM observations if bounds on the tidal deformability and the chirp mass can be inferred from GW astronomy. Although broadly consistent, we obtain a more 
conservative lower bound on the tidal deformability than \citep{RaPe2018}, who find lower bounds of $\tilde{\Lambda} \gtrsim 400$ to form disks and ejecta massive enough to create bright EM observables. On the other hand, the radius constraint derived in~\cite{BaJu2017} is in great agreement with our result, since~\cite{BaJu2017} arrive at $\tilde{\Lambda}>210$. 
Additionally, also a comparison against \cite{AnGo2017} and \cite{MoWe2018} which obtain, respectively, lower bounds on the tidal deformability of 120 and 375 (2
$\sigma$-value) for a 1.4 solar mass NS is possible. \cite{AnGo2017} and \cite{MoWe2018} base their results on constraints obtained from GW170817 and state-of-the-art nuclear physics considerations. While in particular \cite{MoWe2018} obtains a more stringent bound, very similar to the one of \cite{RaPe2018}, this result is in agreement with ours since the bound of \cite{MoWe2018} is based on a large set of possible EOSs and gives credible interval with respect to this comparison set of EOSs and not on the direct measurement of GW170817 or AT2017gfo as done in this work. 
In addition to $q$ and $\tilde{\Lambda}$, 
our analysis also allows us to estimate the amount of dynamical ejecta. 
We find that only 10\% of the total amount of ejecta is dynamical ejecta, 
which supports the idea that the bulk of the ejecta comes from disk outflows \citep{MePi2008}.

\section*{Summary}

In this article, we obtained constraints on the GW170817 progenitor’s 
mass ratio and tidal deformability, which are more stringent than those 
obtained purely from gravitational-wave observations. The unknown equation of state can be constrained once information of the observed GW and EM signals are combined.
To our knowledge, the presented analysis is the first study 
constraining the source properties of GW170817 and EOS with statistical 
methods modeling the lightcurve and spectra of AT2017gfo with surrogate models of radiative transfer simulations, 
see e.g.~\cite{BaJu2017,RaPe2018} for alternative approaches 
combining EM and GW information. 

Concentrating on the lightcurve fits, given that the broadband colors are the most robustly modeled, a 2-component fit is favored over a 1-component fit, although the single-component fit still broadly reproduces the photometric lightcurves well. The single component fit is consistent with a large ejecta mass $M_{\rm ej} \approx 0.05M_{\odot}$ and blue (lanthanide-poor) component ($X_{\rm lan} \approx 3 \times 10^{-4}$). The velocity distribution is broad and slightly bi-modal, partially favoring a low velocity ($v_{\rm ej} \approx 0.06$ $c$) and partially a high ($v_{\rm ej} \approx 0.3$ $c$).

For the two component fit, our findings of a relatively large ejecta mass $M_{\rm ej} \approx 0.03M_{\odot}$, and low velocity $v_{\rm ej} \approx 0.1$ $c$, for the red (lanthanide-rich) component of the ejecta support its origin as being an outflow from the post-merger accretion disk   \citep{MePi2008,FeMe2013,JuBa2015,SiMe2017}, in agreement with previous interpretations of the KN emission from GW170817 (e.g.~\citealt{CoBe2017,ChBe2017,KaMe2017,ViGu2017,RaPe2018}). Three-dimensional MHD simulations imply that $\approx 40\%$ of the newly-formed torus can be ejected in winds at typical speeds $v_{\rm ej} \approx 0.1$ c \citep{SiMe2017}, such that the large inferred ejecta mass for GW170817 is explained by a relatively massive torus, $\approx 0.1\,M_{\odot}$.  GR simulations show that the latter is a fairly generic outcome of the merger process if the merger remnant first goes through a hypermassive NS phase (e.g.~\citealt{ShTa2006}), and thus our observations disfavor a prompt collapse (see also \citealt{MaMe2017,BaJu2017}).  On the other hand, whether the inferred lanthanide mass fraction is sufficient to explain the details of the solar system $r$-process abundance pattern (which requires $X_{\rm lan} \approx 0.03-0.1$) is less clear; our results depend on the assumption of spherical symmetry, which could overestimate the amount of lanthanide-free ejecta. 

By contrast, we infer that the blue (lanthanide-poor) component of the ejecta possesses a somewhat higher velocity $v_{\rm ej} \gtrsim 0.2$ $c$ and a similar ejecta mass $M_{\rm ej} \approx 0.025 M_{\odot}$ than the red component.  While the velocity scale of the blue ejecta naturally matches expectations for the dynamical ejecta (e.g.~\citealt{HoKi13,BaGo2013}), the relatively large quantity that we infer appears in tension with current GR merger simulations which focus on dynamical ejection mechanism.  This may point to an alternative source of blue ejecta, such as the magnetized neutrino-irradiated wind from a long-lived hypermassive NS remnant prior to its collapse to a black hole (\citealt{MeTh2018}; a purely neutrino-driven outflow is insufficient to explain the observed properties; \citealt{DeOt2008}).  Alternatively, as with the red ejecta, the blue ejecta could originate from an accretion disk outflow (e.g.~\citealt{MeFe2014,PeRo2014}); however, the high velocity is incompatible with both hydrodynamical and MHD simulations (e.g.~\citealt{FaMe2014,SiMe2017}).  

Some of the blue light seen at the earliest epoch $\lesssim 1$ day could in principle also be attributed to physical effects not included in our modeling, such as the decay of free neutrons in the outermost fastest parts of the ejecta \citep{Ku2005,MeBa2015}, or additional thermal energy added to the ejecta by a relativistic jet (``cocoon'' emission; \citep{GoNa2017,KaNa2017,PiKo2018}, however, see \cite{DuQu2018}, who find that relatively little thermal energy is imparted to the ejecta to power early blue emission in the case of a successful gamma-ray burst jet) or by internal shocks within whatever variable and temporally-extended source (magnetar wind or accretion disk outflow) produces the KN ejecta \citep{MeTh2018}.  As already discussed, we cannot exclude that up to $\sim 10\%$ of the ejecta ($\lesssim 6\times 10^{-3}M_{\odot}$) is dynamical in origin and instead could originate, e.g.~from the tidal tail.  The tidal tail ejecta is predicted to be fast ($v_{\rm ej} \approx 0.2-0.3$ c) and lanthanide-rich ($X_{\rm lan} \gtrsim 0.03$), and its contribution to the light curve may be swamped by other components in the case of NS-NS mergers; prospects are better for unambiguously detecting this component in a NS-BH merger (e.g.~\citealt{FoDe2017}).
 
Further work is needed due to possible systematic 
uncertainties introduced by the 
computation of the ejecta mass in numerical relativity 
simulations~\citep{DiUj2017,CoDi2017,AbEA2017f} 
and the assumptions of our lightcurves as the restriction to spherical geometry.
Since opacity and velocity control the diffusion time of the ejecta, 
the different ejecta channels have different characteristic magnitude, 
color, and durations. This is further complicated by the fact that the
observed color is viewing angle dependent \citep{KaFe2015} and 
that dynamical ejecta can have a gravitationally bound component 
falling back onto the central object, interacting with the outflow and altering 
the mass ejection and composition of the disk. 
\cite{FeQu2015} showed that the disk outflow suppresses fallback accretion, 
and \cite{FeFo2017} extended this analysis by varying the relative mass ratios 
of the ejecta by changing the density of the dynamical ejecta.
Furthermore, the accuracy of current radiative transfer models in predicting kilonova colors still needs to be fully investigated, and more work is needed to improve atomic line lists, 
transfer physics, and thermalization \citep{KaBa2013,MaKa2017,Ta2016}. 
Nevertheless, we have shown how the lightcurve and spectra 
can be robustly modeled and how parameter estimation 
pipelines can be employed to determine the source properties
from the EM observations. \\

The lightcurves used in this analysis are publicly available at:  
\url{https://github.com/dnkasen/Kasen\_Kilonova\_Models\_2017}.
The lightcurve fitting code is available at: 
\url{https://github.com/mcoughlin/gwemlightcurves}.

\acknowledgments

MC is supported by the David and Ellen Lee Postdoctoral Fellowship at the California Institute of Technology.
TD acknowledges support by the European Union’s Horizon 2020 research and 
innovation program under grant agreement No 749145, BNSmergers.  ZD is supported by NSF Graduate Research
Fellowship grant DGE-1144082. SJS acknowledges funding from STFC grant ST/P000312/1. 
AJ acknowledges funding by the European Union’s Framework Program for Research and Innovation Horizon 2020 under Marie Sklodowska-Curie grant agreement No. 702538.
GL is supported by a research grant (19054) from VILLUM FONDEN.
ROS is supported by NSF award PHY-1707965. 

\bibliographystyle{aasjournal}
\bibliography{references}

\appendix

\section{Lightcurves}
\label{sec:lightcurves}

There were some differences in the bolometric luminosity estimated by authors when the first data papers were published on AT2017gfo \citep[see for example][]{2017Sci...358.1570D,SmCh2017} particularly after 9-10 days.  \cite{WaOf2017} have compiled the published data and calculated a
bolometric lightcurve and we employ a similar approach here. For example \cite{SmCh2017} only integrated between the observed filter ranges and did not extrapolate beyond 2.5$\mu$m whereas 
\cite{2017Sci...358.1570D} and \cite{WaOf2017} used either a blackbody extrapolation or
power law $f(\lambda)\sim \lambda^{-4}$. The real level of emitted flux beyond 2.5$\mu$m 
is not constrained by any data available and remains a significant unknown. 

We furthermore 
reconstructed the bolometric light curve for AT2017gfo between +0.47\,d and +10.4\,d based on these 20 distinct
epochs of photometry. In most cases we had sufficient broad band fluxes to 
construct a black body fit. However, in some cases, we opted to use interpolated photometry
to have a data point for missing bands. For example in order to make best use of the early Swift photometry, we have extrapolated ground based optical photometry, but we consider the resulting fit very reasonable. We then fit Planck functions to the multi-wavelength photometry, assuming that the emission can be described by a blackbody. We thus determined the black body temperature and radius and their 68\% uncertainties by use of the MATLAB function \emph{fit}. 
Our estimated bolometric luminosity is therefore the total luminosity emitted at all wavelengths of a black body emitter of that temperature and radius.
We have used Monte Carlo resampling to determine the asymmetric errors of the bolometric luminosity. In general, a single blackbody fits satisfactorily up to 5.4 days. It starts to deviate at 6.4, getting worse through 7.4 and 8.4 days and by day 9.4 onwards it is clearly not a single blackbody SED. 
We therefore consider the bolometric properties determined with this method unreliable past this point in time.  Our revised bolometric light curve is provided in Table~\ref{table:bolometric}.
We have compared our bolometric light curve with those presented by \cite{WaOf2017} and \cite{Arc2018}, and we find good agreement in the central values. However, we consider our uncertainties more reasonable, given the assumptions and the photometry errors, while a few points in \cite{WaOf2017} have unrealistically small uncertainties (below 3\%).  Our updated $L_{\rm bol}$ values differ from those in \cite{SmCh2017} in particular as that paper only integrated out to the spectral energy distribution out to the red edge of the $K$-band filter. 

\begin{table}
\centering
\begin{tabular}{rrrr}
\toprule
Phase [days] & L [erg/s] & $\sigma$ L$^-$ [erg/s] & $\sigma$ L$^+$ [erg/s] \\
\bottomrule
        0.47 &   9.4e+41 &            2.8e+41 &            3.7e+41 \\
        0.64 &   8.6e+41 &            3.7e+41 &            5.5e+41 \\
        0.70 &   8.6e+41 &            6.2e+41 &            1.5e+42 \\
        0.88 &   9.8e+41 &            4.0e+41 &            5.7e+41 \\
        1.05 &   5.1e+41 &            2.1e+41 &            3.0e+41 \\
        1.22 &   4.3e+41 &            8.1e+40 &            9.5e+40 \\
        1.43 &   3.8e+41 &            1.1e+41 &            1.3e+41 \\
        1.69 &   2.6e+41 &            2.2e+41 &            7.8e+41 \\
        1.82 &   1.8e+41 &            1.6e+41 &            7.8e+41 \\
        2.21 &   2.6e+41 &            5.3e+40 &            6.6e+40 \\
        2.42 &   2.3e+41 &            7.2e+40 &            9.2e+40 \\
        2.68 &   1.5e+41 &            1.3e+41 &            5.6e+41 \\
        2.83 &   1.8e+41 &            8.3e+40 &            1.3e+41 \\
        3.22 &   2.1e+41 &            3.0e+40 &            3.6e+40 \\
        3.41 &   1.5e+41 &            5.0e+40 &            6.8e+40 \\
        4.14 &   1.5e+41 &            7.5e+40 &            1.2e+41 \\
        4.40 &   1.5e+41 &            4.2e+40 &            5.5e+40 \\
        5.40 &   1.1e+41 &            4.5e+40 &            6.6e+40 \\
        6.40 &   1.1e+41 &            2.7e+40 &            3.3e+40 \\
        7.40 &   6.6e+40 &            1.7e+40 &            2.0e+40 \\
        8.40 &   3.6e+40 &            1.5e+40 &            2.1e+40 \\
        9.40 &   1.7e+40 &            7.6e+39 &            1.2e+40 \\
       10.40 &   5.2e+39 &            4.5e+39 &            1.6e+40 \\
       11.30 &   1.2e+40 &            1.0e+40 &            2.8e+40 \\
       13.21 &   6.8e+39 &            6.7e+39 &            1.0e+41 \\
\bottomrule
\end{tabular}
\caption{Bolometric lightcurve values (and error bars) used in the analysis.
}
\label{table:bolometric}
\end{table}

\section{Surrogate Model}
\label{sec:surrogate}

The prescription for the algorithm is as follows. 
First, each bolometric lightcurve, photometric lightcurve, and spectral energy distribution in the simulation set is sparsely interpolated onto the same time array of 0.1\,days, which is more densely sampled than most of the data.
The bolometric lightcurve and the photometric lightcurves in the various passbands are computed directly from the spectra.
For the photometric lightcurves, each passband is analyzed separately and for the spectra, each wavelength is analyzed separately.  
We denote these vectors of photometry or spectra in one frequency bin for different times as $\tau_{i}(M_{\rm ej}^j,v_{\rm ej}^j,X_{\rm lan}^j)$ (where $i$ is the $i$-th time and $j$ is the $j$-th set of ejecta parameters on the simulation grid) and the matrix of such vectors as $\mathcal{T}_{ij} = [\tau_{i}(M_{\rm ej}^j,v_{\rm ej}^j,X_{\rm lan}^j)]$. Rather than interpolate the $i$-th component of $\tau_i$ as a function of $(M_{\rm ej},v_{\rm ej},X_{\rm lan})$, we instead interpolate principal components of each $\tau_i$ vector since entries of $\tau_i$ co-vary\footnote{For simplicity, we ignore the covariance between different frequency bins, which may be included in future analyses.}.     
Performing a singular value decomposition (SVD) of this matrix
\begin{equation}
\mathcal{T} = V\Sigma U^\top
\end{equation}
yields orthonormal basis vectors in the columns and rows of $V$ and $U$.  We then project each $\tau_{i}$ into the left-singular vector basis
\begin{equation}
s_{k}(M_{\rm ej}^j,v_{\rm ej}^j,X_{\rm lan}^j) = V_{ki}^\top \tau_{i}(M_{\rm ej}^j,v_{\rm ej}^j,X_{\rm lan}^j)
\end{equation}
using all available basis vectors\footnote{The basis is often truncated in many applications to minimize computational resources, but here we keep all basis vectors.}. (Note that Einstein summation notation is used above).  This projection results in the $s_k$ components being weights of principal components of the input data $\mathcal{T}_{ij}$.  

We now independently interpolate the $k$-th component of $s_{k}$, conditioning on the known $s_{k}(M_{\rm ej}^j,v_{\rm ej}^j,X_{\rm lan}^j)$.  The interpolation is done using Gaussian process regression \citep[GPR,][]{GPRBook}, a statistical interpolation method which produces a posterior distribution on a function $f$ given known values of $f$ at a few points in the parameter space.
Here we describe the basic formulation and facets of Gaussian process regression.  We refer the interested reader to \cite{GPRBook} for a comprehensive and pedagogical description of Gaussian processes.  The essential assumption in GPR is that neighboring values of a function $f(\vec{\theta})$ and $f(\vec{\theta}')$ are correlated, and that their joint distribution is a multivariate Gaussian fully described by a mean and covariance. The covariance between function values is prescribed in a kernel function $k(\vec{\theta},\vec{\theta}')$ that typically depends only on the distance between points $\vec{\theta}$ and $\vec{\theta}'$. A common choice is a Gaussian kernel, for example.  To perform a regression, function values ${\bf f_*}$ at points $\Theta_*$ are inferred by conditioning on known function values after choosing a kernel function.  The parameters and/or form of the kernel (called {\it hyperparameters}), e.~g.~the Gaussian width, are usually optimized to maximize the evidence for known ${\bf f}$ values. Following \cite{GPRBook} and assuming a zero-mean prior, the posterior distribution on function values ${\bf f}_*$ at points $\Theta_*$ conditioned on known values ${\bf f}$ at $\Theta$ has a mean given by
\begin{equation}
K(\Theta_*,\Theta)K(\Theta,\Theta)^{-1}{\bf f}
\end{equation}
and covariance
\begin{equation}
K(\Theta_*,\Theta_*) - K(\Theta_*,\Theta)K(\Theta,\Theta)^{-1}K(\Theta,\Theta_*)
\end{equation}
where the $K$ matrices are the covariance matrices between known and/or inferred function values computed from the kernel.  The mean can be used as a simple interpolator, or the full posterior distribution can be used if samples or uncertainties are of interest.

Specifically, we employ the \texttt{sci-kit learn} implementation of GPR \citep{scikit-learn}. Before interpolation, each $s_k$ is \textit{whitened}:
\begin{equation}
s_k^{\rm whitened}(M_{\rm ej}^j,v_{\rm ej}^j,X_{\rm lan}^j) 
=\frac{s_{k}(M_{\rm ej}^j,v_{\rm ej}^j,X_{\rm lan}^j) - \underset{j}{\rm mean}\left( s_{k}(M_{\rm ej}^j,v_{\rm ej}^j,X_{\rm lan}^j)\right)}{\underset{j}{\rm range}\left(s_{k}(M_{\rm ej}^j,v_{\rm ej}^j,X_{\rm lan}^j)\right)}. 
\end{equation}
where ``range'' indicates the difference of the maximum and minimum values.
The mean value of $s_k^{\rm whitened}$ for arbitrary  $(M_{\rm ej},v_{\rm ej},X_{\rm lan})$ is then regressed with a zero-mean Gaussian process conditioned on $s_k^{\rm whitened}(M_{\rm ej}^j,v_{\rm ej}^j,X_{\rm lan}^j)$.  We assume a rational-quadratic kernel function of the form
\begin{equation}
k(\vec{\theta},\vec{\theta}') = \left(1+\frac{|\vec{\theta}-\vec{\theta}'|^2}{2\alpha l^2}\right)^{-\alpha}
\end{equation}
where $\vec{\theta}$ and $\vec{\theta}'$ are vectors of input parameters $(M_{\rm ej},v_{\rm ej},X_{\rm lan})$.  The hyperparameters $\alpha$ and $l$ are chosen by maximizing the evidence for the data under a zero-mean Gaussian process.  

The interpolated $s_k^{\rm whitened}(M_{\rm ej},v_{\rm ej},X_{\rm lan})$ is then de-whitened and projected back into the time domain:
\begin{equation}
\tau_i(M_{\rm ej},v_{\rm ej},X_{\rm lan}) = V_{ik} s_k(M_{\rm ej},v_{\rm ej},X_{\rm lan})
\end{equation}

\begin{figure*}[t] 
\begin{center}
 \includegraphics[width=3.5in]{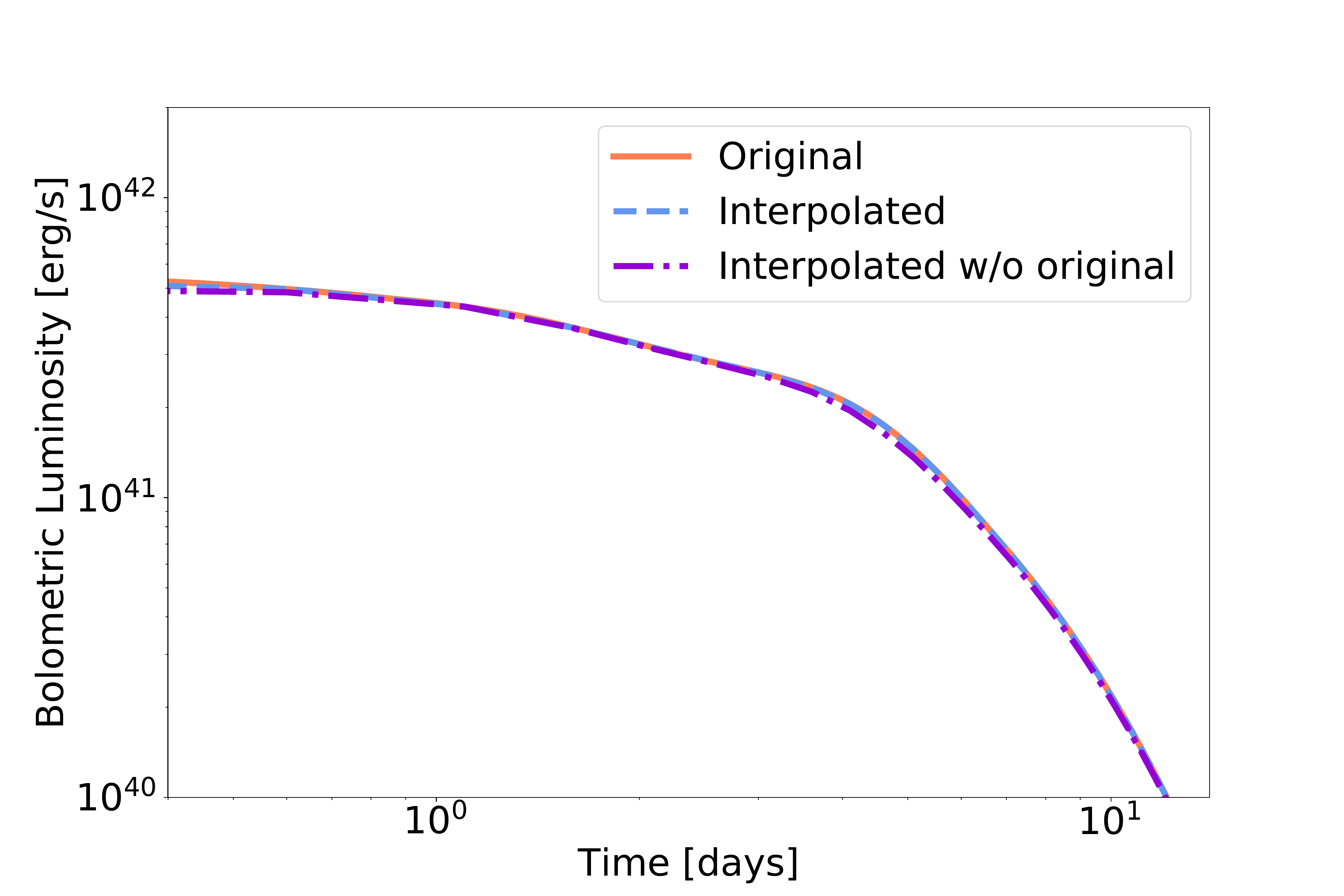}
\end{center}
 \includegraphics[width=3.5in]{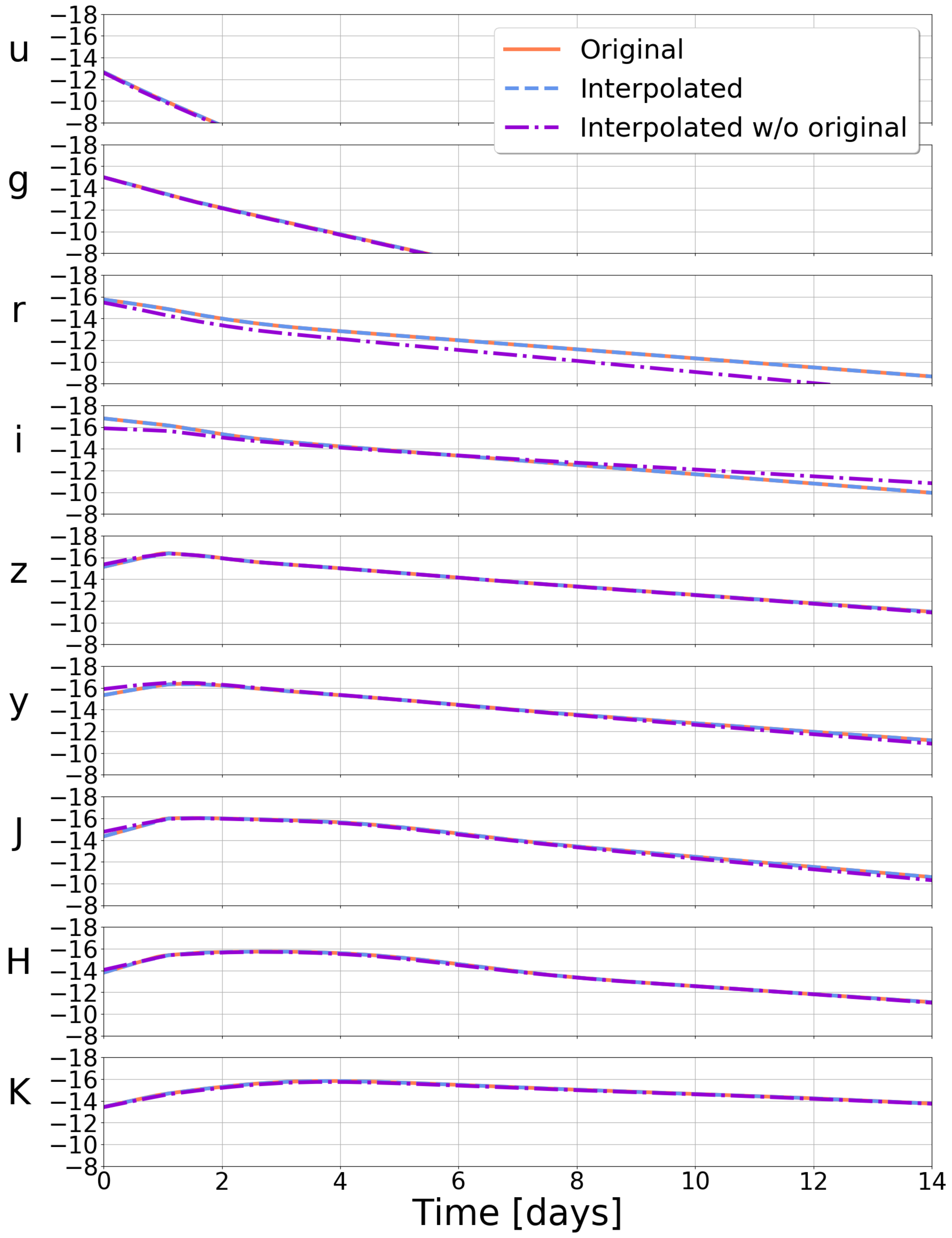}
\includegraphics[width=3.5in]{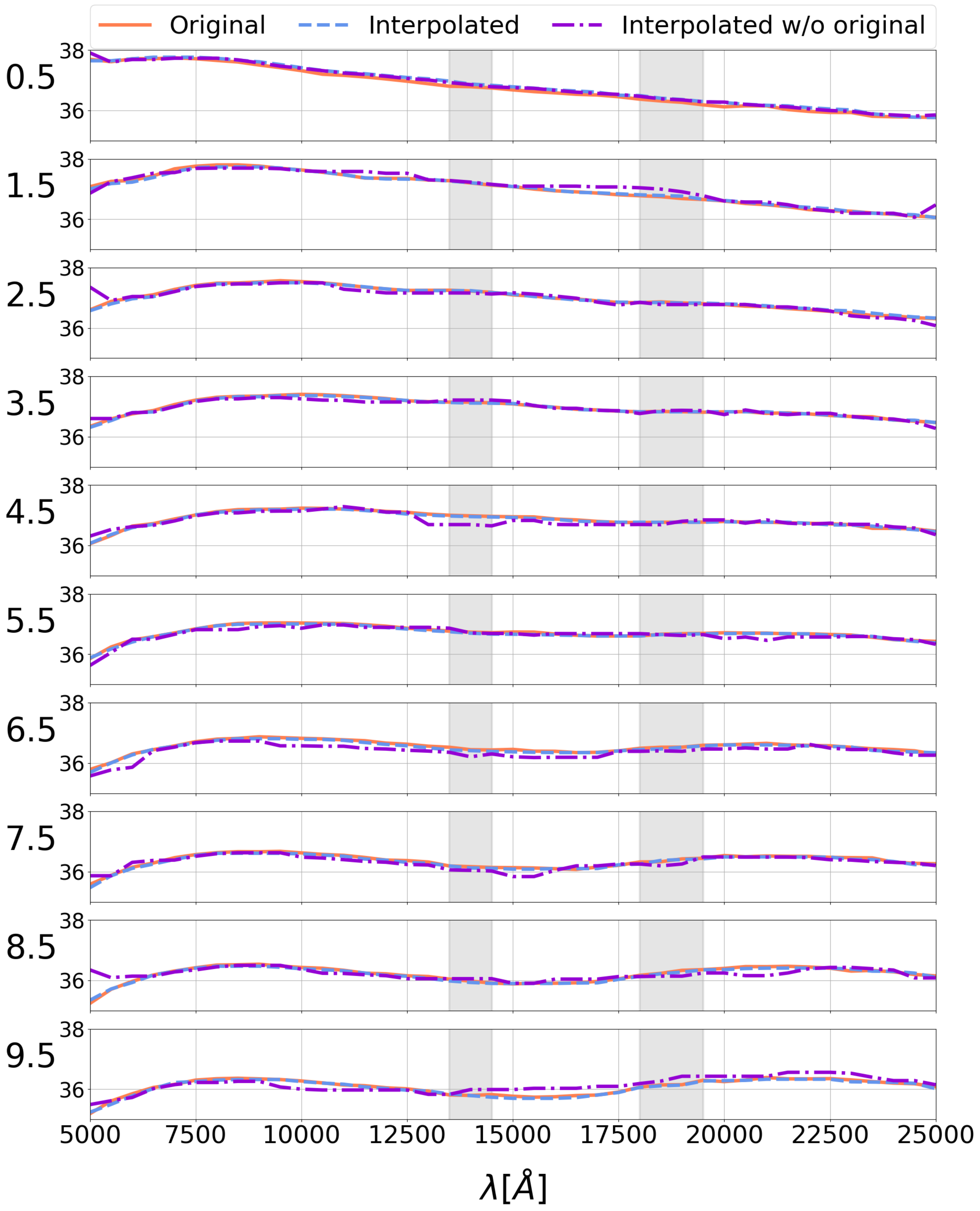}

  \caption{
     Comparison of original bolometric luminosity (top), lightcurves (bottom left), 
     and spectra (bottom right) with the interpolated model using all points on the grid and an interpolated model with the grid point being tested removed. 
     The grid point tested is $X_{\rm lan} = 0.001$, $M_{\rm ej} = 0.05$
and $v_{\rm ej} = 0.2$. }
 \label{fig:validate}
\end{figure*}

The interpolated $\tau_i(M_{\rm ej},v_{\rm ej},X_{\rm lan})$ is used in computation of the likelihood in the Bayesian inference presented in the next section.  The GPR mean is only used here, but future work will incorporate uncertainties from the GPR. 
We seek to validate the interpolated model using the standard technique of removing the model interpolated at a point ($X_{\rm lan} = 0.001$, $M_{\rm ej} = 0.05$, and $v_{\rm ej} = 0.2$) and comparing the model both with and without its inclusion. Figure~\ref{fig:validate} shows a comparison of original bolometric luminosity (bottom left), lightcurves (upper left), and spectra (upper right) at this point. The model without the missing point is nearly indistinguishable across the examples here, while the model with the missing point is within error bars of 1\,mag assumed in the analysis.

\section{Corner Plots}
\label{sec:corner}

\begin{figure*}[t] 
 \includegraphics[width=3.5in,height=3in]{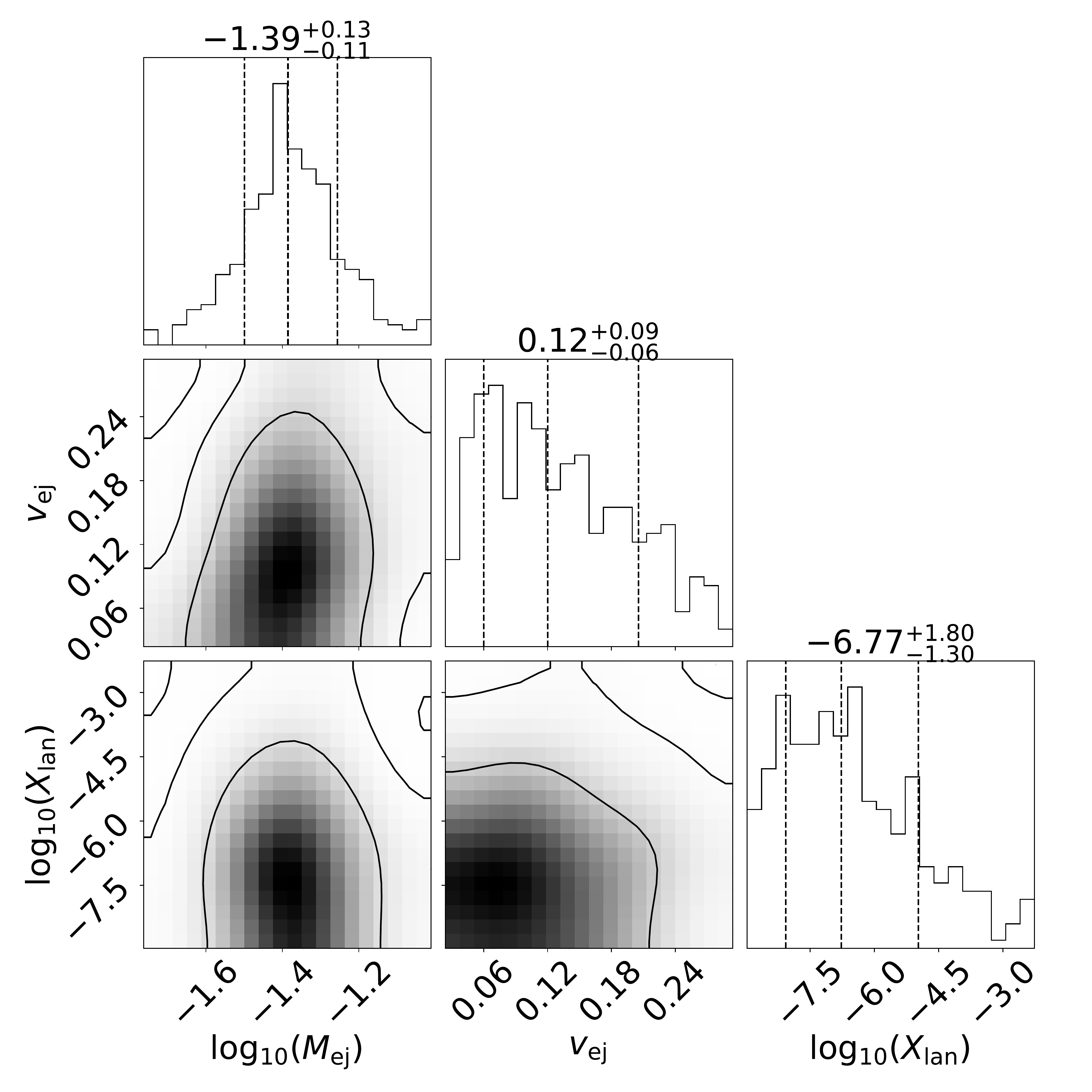}
  \includegraphics[width=3.5in,height=3in]{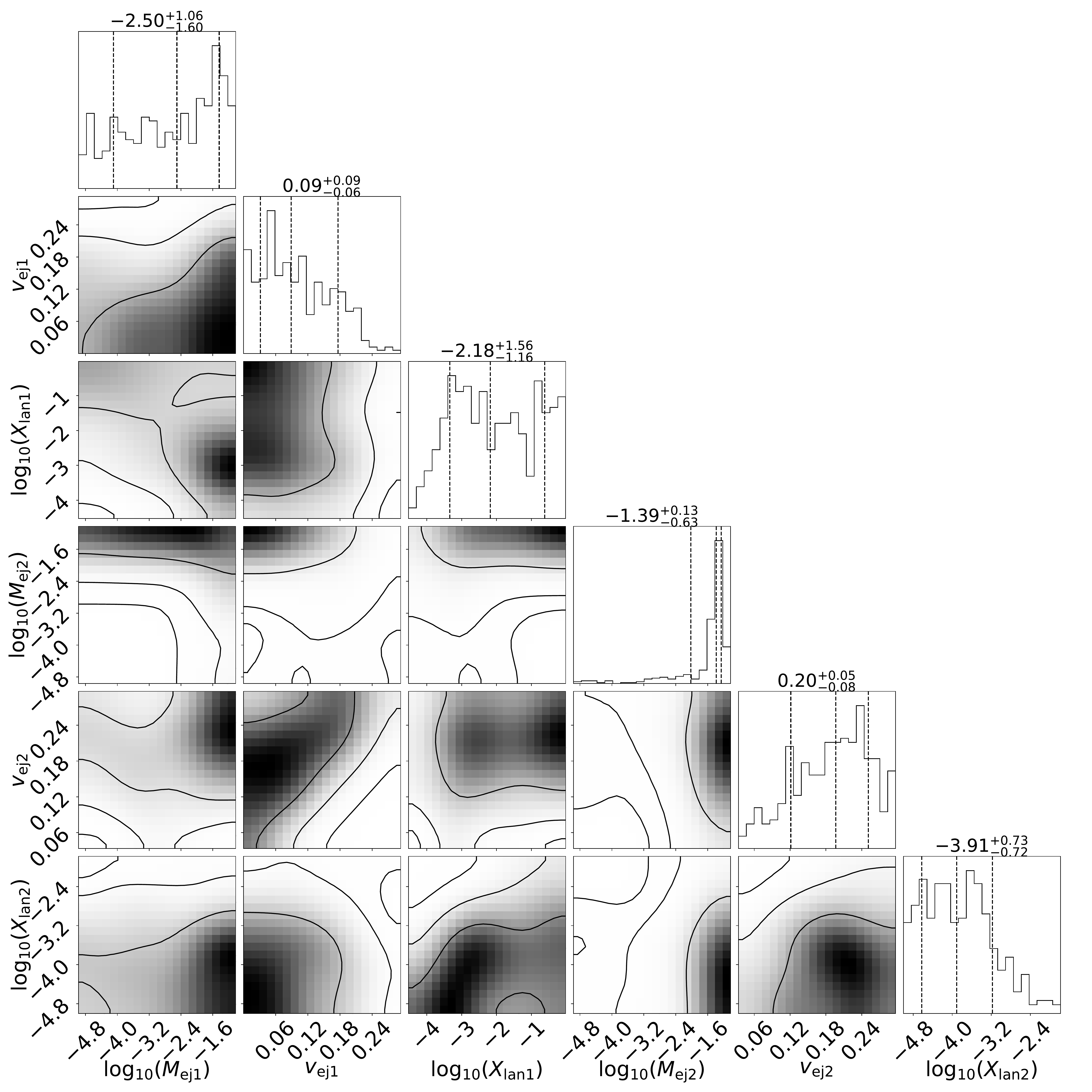}
\includegraphics[width=3.5in,height=3in]{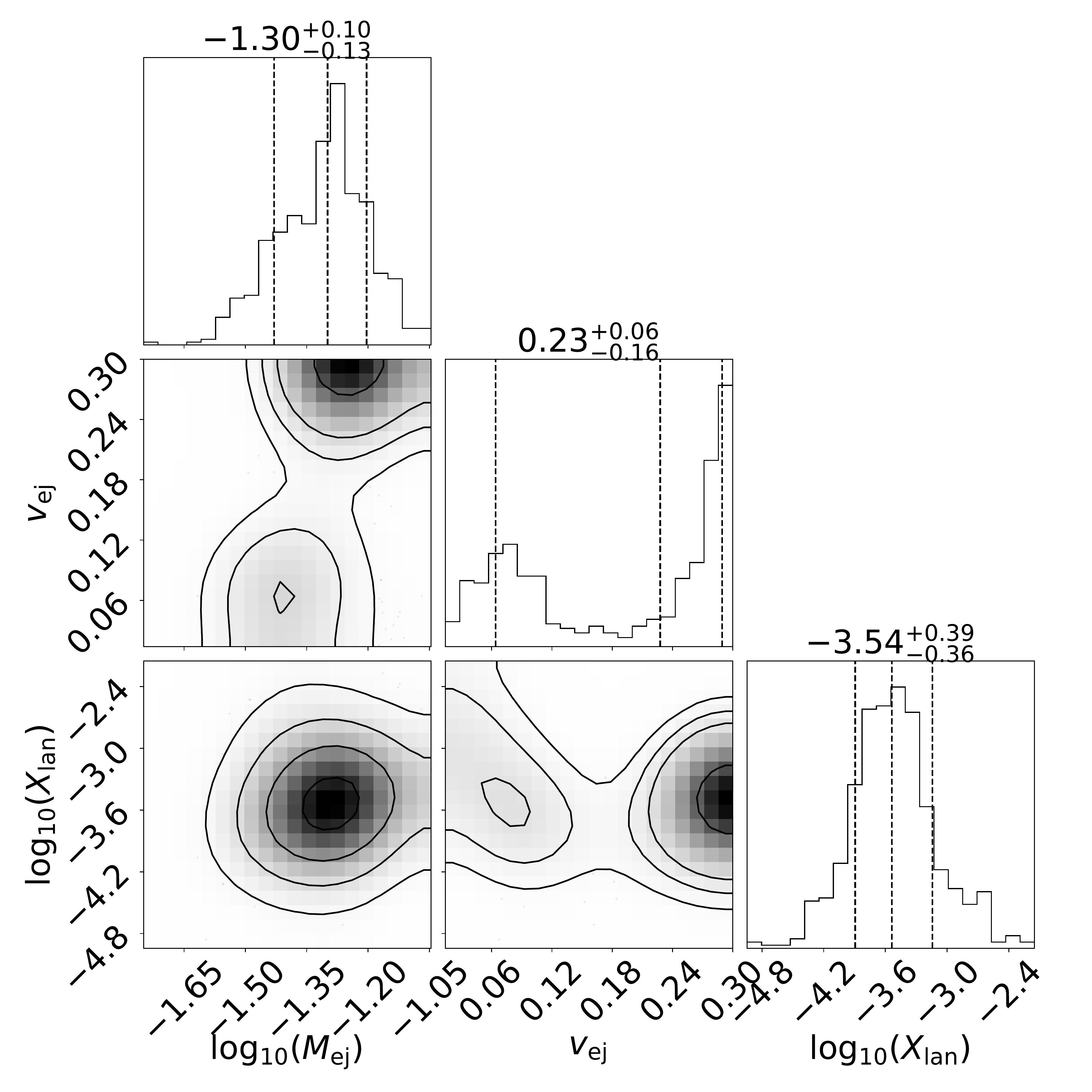}
\includegraphics[width=3.5in,height=3in]{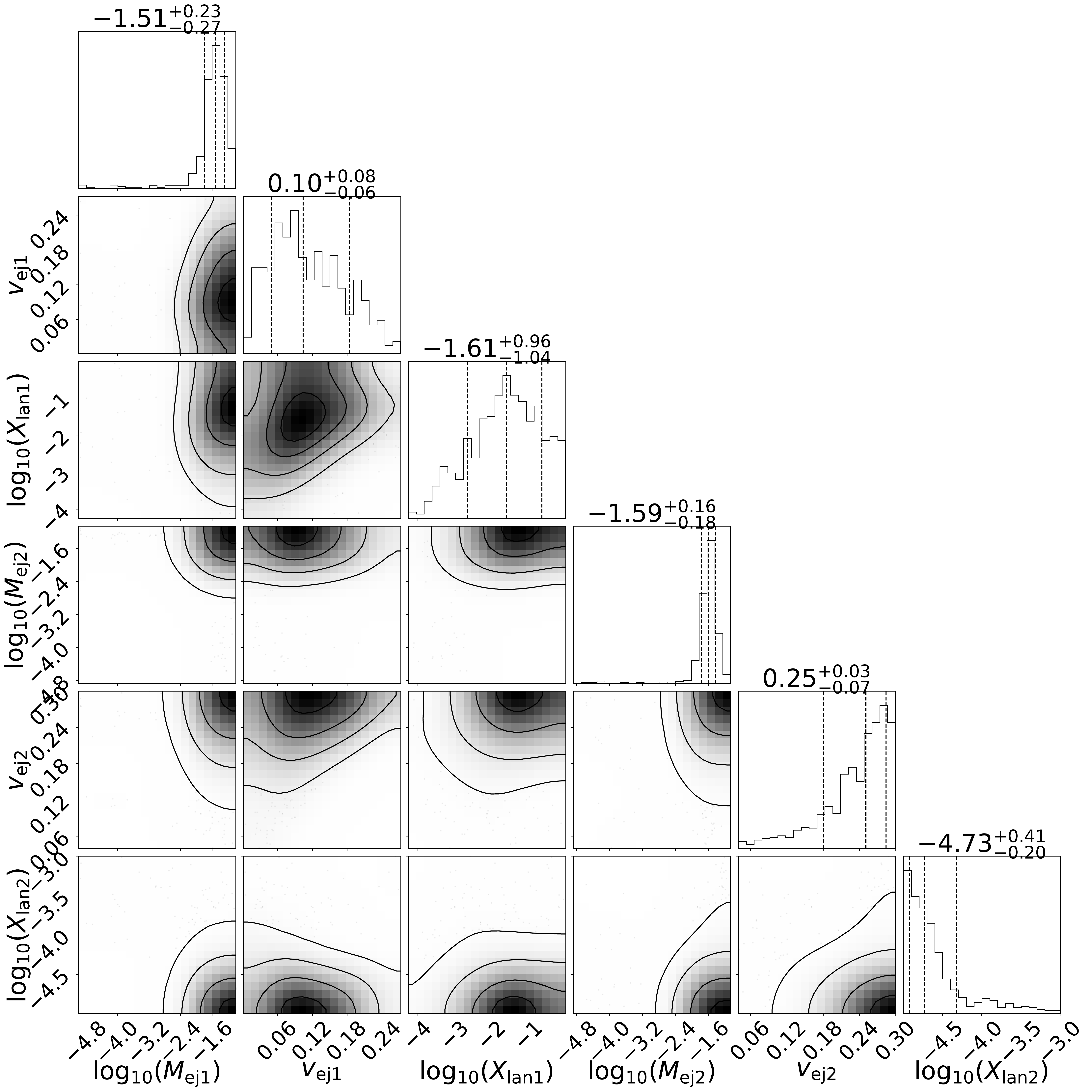}   
\includegraphics[width=3.5in,height=3in]{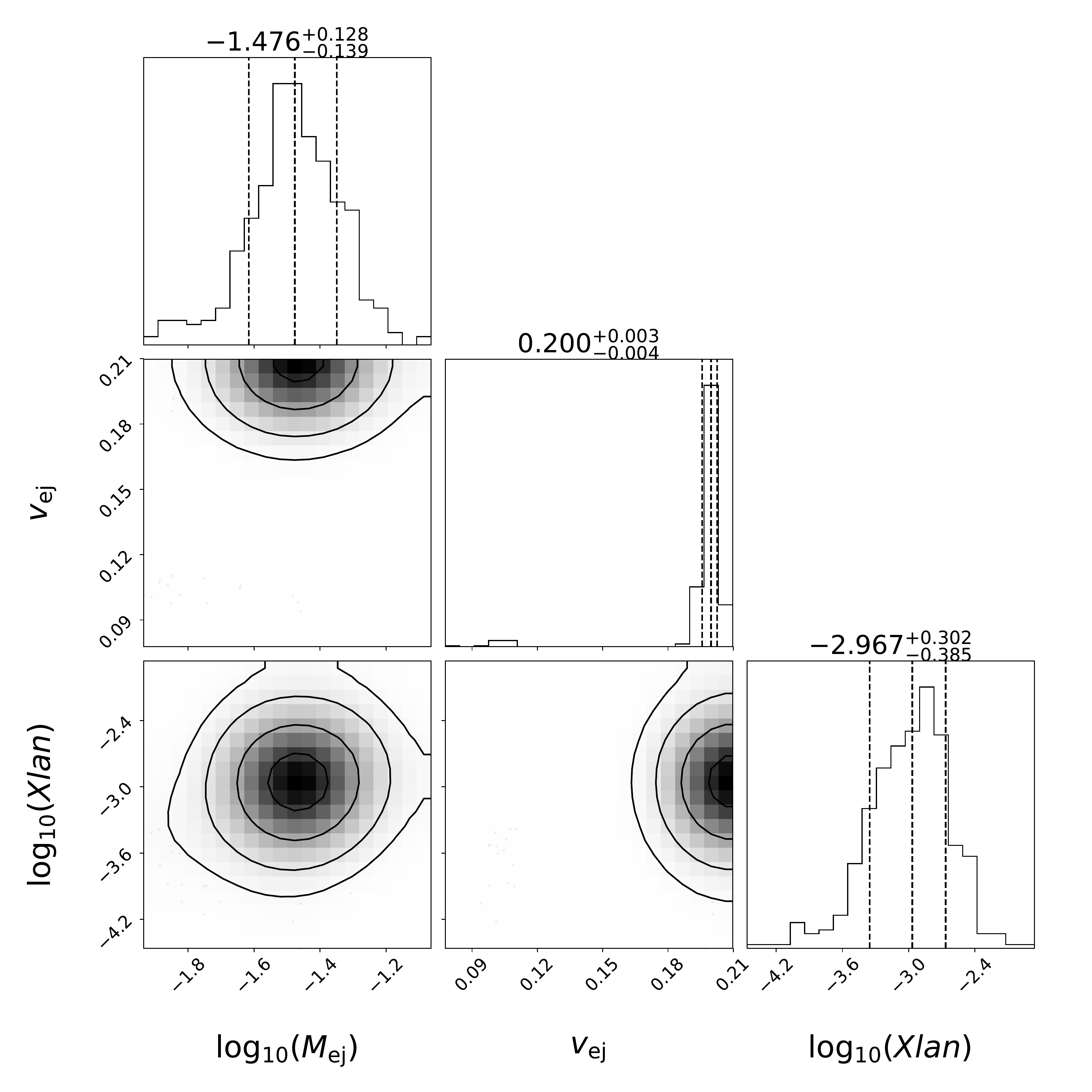}
\includegraphics[width=3.5in,height=3in]{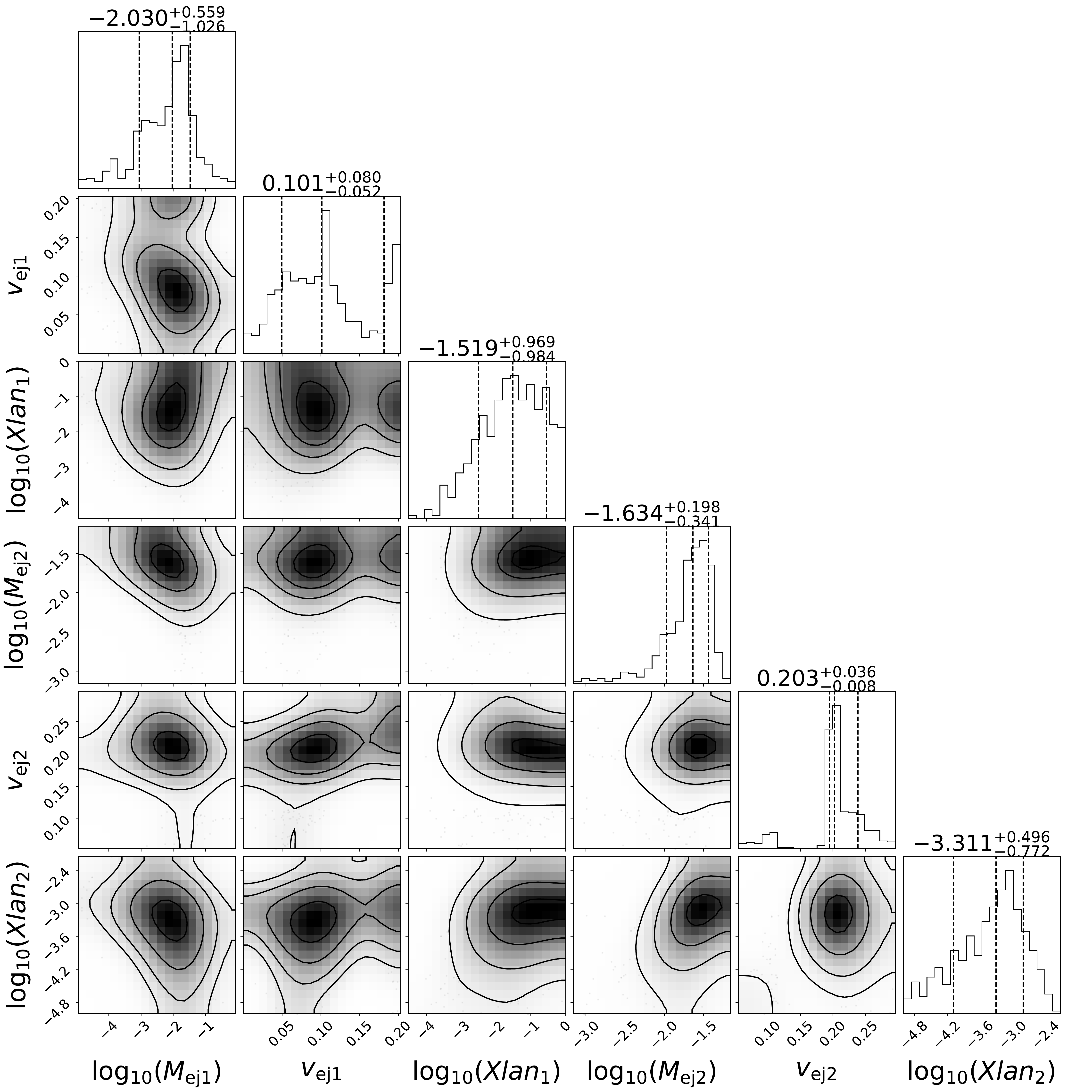} 
  \caption{
     The corner plots for both one (left-column) and two (right-column) component models from \cite{KaMe2017}, for the bolometric luminosity (top row), lightcurve (middle row), and spectra (bottom row).
        The source of the photometry is summarized in section~\ref{sec:data}.
       X-shooter spectra is compiled from \cite{PiDa2017} and \cite{SmCh2017}.}
 \label{fig:corner}
\end{figure*}

Figure~\ref{fig:corner} shows the associated ``corner'' plots \citep{For2016}, quantifying the level of overlap between parameters using 1- and 2-D posteriors marginalized over the rest of the parameters.

\section{Fits based on the lightcurves and spectra}
\label{sec:fits}

\begin{figure*}[t]
 \includegraphics[width=3.5in]{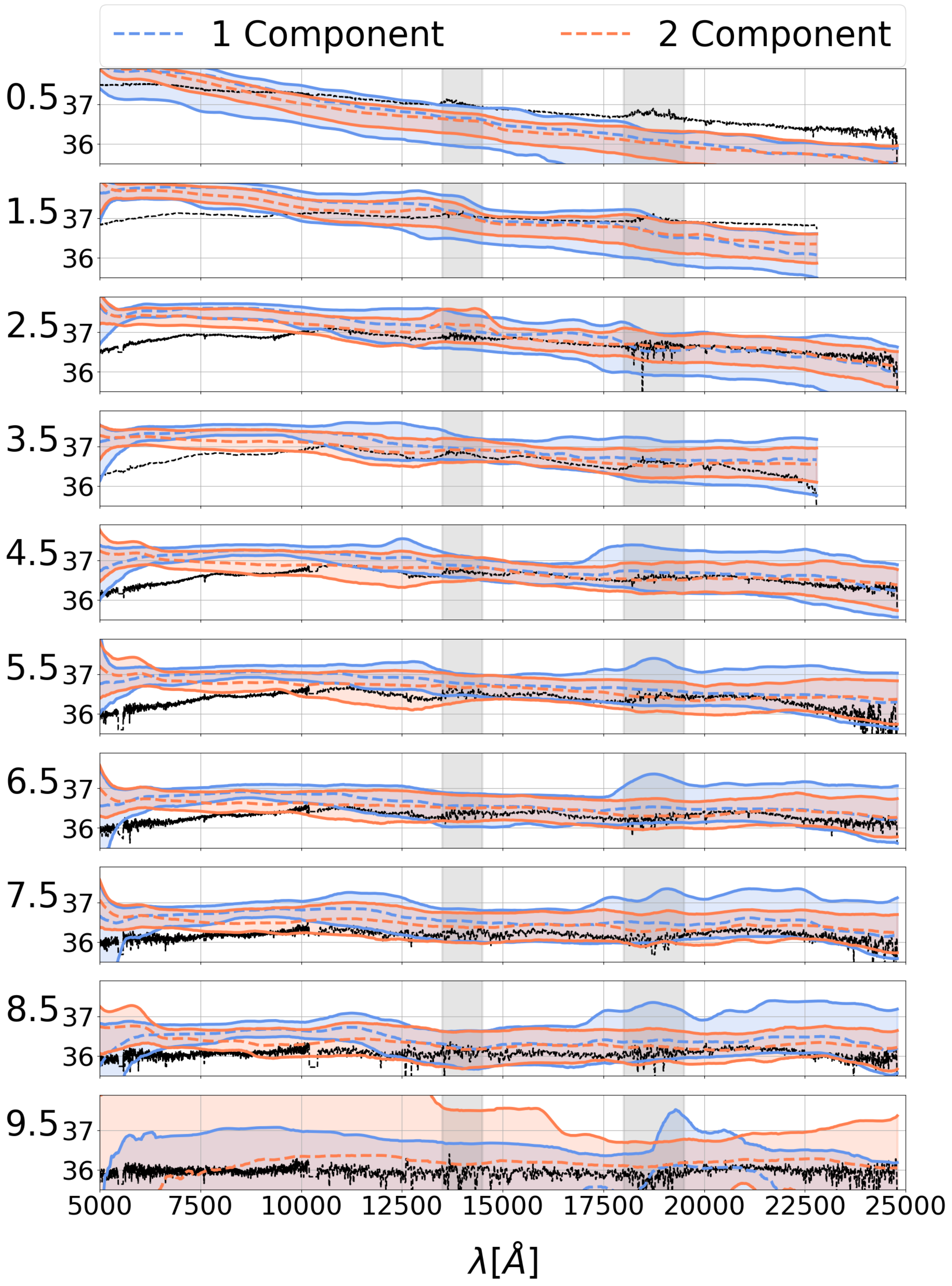}
  \includegraphics[width=3.5in]{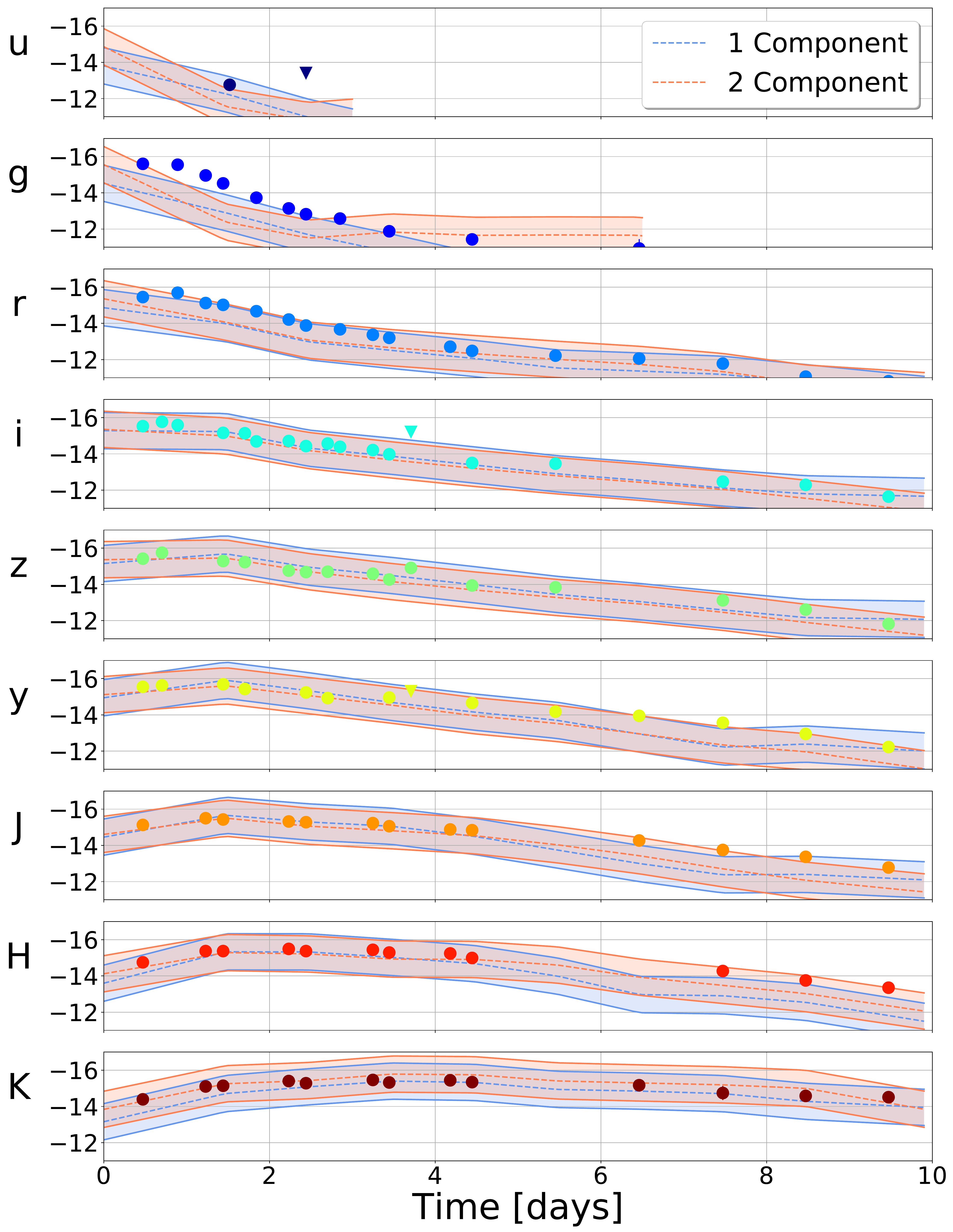}
 \caption{
    X-shooter spectra (black lines) at the available epochs and one and two component model fits from the lightcurve analysis \citep{PiDa2017,SmCh2017}.
    The dashed lines show the median spectrum, while the shaded intervals show the 90\% intervals.
    The numbers to the left of the y-axis show the approximate epochs of the observations.
    The gray vertical shaded regions correspond to parts of the spectrum contaminated by atmospheric transmission.
    On the right are the photometry with lightcurves derived from the spectra fits.
  }
  \label{fig:fits}
\end{figure*}

Figure~\ref{fig:fits} shows the spectra based on the lightcurve fits (and vice-versa).
As explained in the main text we find consistency between fits 
obtained from the lightcurves or spectra directly.

\section{Numerical Relativity Fits}
\label{sec:NR}

In this article, we improve the fits of~\cite{DiUj2017}
to obtain better constraints on the source properties. 
The two main improvements are that we include a larger set of 
numerical relativity simulations using results presented in~\cite{DiUj2016,HoKi13,DiBe2015,BaGo2013,LeLi2016,SeKi2016,BoMa2017,ShFu2017,CiKa2017} and that we fit 
$\log_{10}(M_{\rm ej})$ instead of $M_{\rm ej}$. We obtain
\begin{small}
\begin{equation}
\log_{10} (M_{\rm ej}^{\rm NR}) = \left[ \frac{a (1-2 C_1) M_1}{C_1}+b M_2 \left(\frac{M_1}{M_2}\right)^n  + \frac{d}{2}\right]  +[1 \leftrightarrow 2] \label{eq:Mej}
\end{equation}
\end{small}
where $[1 \leftrightarrow 2]$ indicates the sum is repeated with indices switched, with $a=-0.0812,b=0.2288,d=-2.16,n=-2.51$ and
\begin{equation}
 v_{\rm ej}^{\rm NR} = \left[ \frac{e M_1 (f C_1 +1)}{M_2}+ \frac{g}{2}\right] +[1 \leftrightarrow 2] \label{eq:vej}
\end{equation}
with $e = -0.3292, f = -1.633, g = 0.720$, 
where $M_{1,2},C_{1,2}$ denote the mass and compactness of the individual stars.

To obtain constraints on the supranuclear equation of state, 
we rewrite Eqs.~(\ref{eq:Mej}) and (\ref{eq:vej}) 
to be a function of the tidal deformability 
\begin{equation}
 \tilde{\Lambda} = \frac{16}{13} \left[ \frac{M_1+12M_2}{(M_1+M_2)^5}M_1^4 \tilde{\Lambda}_1 \right]+ [1\leftrightarrow 2].  \label{eq:lambda}
\end{equation}
where $[1 \leftrightarrow 2]$ indicates the sum is repeated with indices switched and $\tilde{\Lambda}_1$ and $\tilde{\Lambda}_2$ 
are the tidal parameters associated with the individual NSs, 
and by employing the quasi-universal relations of \cite{Yagi:2016bkt} and 
assuming that $M_1/M_2=C_1/C_2$ which 
is a valid approximation for realistic masses and compactnesses. 
We sample uniformly in $q$ and $\tilde{\Lambda}_1$, which uniquely determines $\tilde{\Lambda}$.
The constraints on $q$, $\tilde{\Lambda}$, and $A$ are driven by their predictions for $M_{\rm ej}$ and $v_{\rm ej}$, in comparison with the measured values from the lightcurve analysis.
In general, as either $q$ or $\tilde{\Lambda}$ increases, $M_{\rm ej}$ increases as well.

Note that within our analysis presented in the main text, we do not place 
any constraints on $q$ from the gravitational-wave analysis.
In principle, the posteriors from the gravitational-wave analysis or binary neutron-star population studies could be used to further constrain the distribution 
of mass ratio or $\tilde{\Lambda}$, but we choose not to do so here.
We only impose $M_c=1.188M_{\odot}$ and $\tilde{\Lambda} \lesssim$ 640 and then employ 
employ Eqs.~(\ref{eq:Mej},\ref{eq:vej}) to determine the mass ratio 
and tidal deformability of the system and with Eq.~\eqref{eq:AMej} to understand how much 
mass is ejected due to dynamical ejecta mechanism.

\end{document}